\documentclass[a4paper,twocolumn,10pt]{article}
\usepackage[utf8]{inputenc}
\usepackage{amsmath}
\usepackage{authblk}
\usepackage[caption=false, position=top, singlelinecheck=off, justification=raggedright]{subfig}
\usepackage[separate-uncertainty=true,binary-units]{siunitx}
\usepackage{booktabs}
\usepackage{graphicx}
\usepackage{tikz}
\usepackage{float}
\usepackage{glossaries}
\usepackage{geometry}
\usepackage[capitalize]{cleveref}
\usepackage{dsfont}
\usepackage[numbers,sort&compress]{natbib}
\usepackage[font=small]{caption}
\usepackage{romannum}

\DeclareMathOperator*{\argmin}{arg\,min}

\definecolor{nicegray}{HTML}{707070} 
\definecolor{nicered}{HTML}{AF5A50}
\definecolor{niceblue}{HTML}{005B82}
\definecolor{nicegreen}{HTML}{7D966E}
\definecolor{niceyellow}{HTML}{D7AA50}

\newacronym{stdp}{STDP}{spike-timing dependent plasticity}
\newacronym{lif}{LIF}{leaky integrate-and-fire}
\newacronym{pid}{PID}{partial information decomposition}
\newacronym{h}{H}{entropy}
\newacronym{te}{TE}{transfer entropy}
\newacronym{ais}{AIS}{active information storage}
\newacronym{i}{I}{mutual information}
\newacronym{mc}{MC}{memory capacity}

\title{Control of criticality and computation in spiking neuromorphic networks with plasticity}
\author[1]{Benjamin Cramer}
\author[1]{David St\"ockel}
\author[1]{Markus Kreft}
\author[2]{Michael Wibral}
\author[1]{Johannes Schemmel}
\author[1]{Karlheinz Meier}
\author[3,4,5]{Viola Priesemann}
\affil[1]{Kirchhoff-Institute for Physics, Heidelberg University}
\affil[2]{Campus Institute for Dynamics of Biological Networks, Georg-August University, G\"ottingen}
\affil[3]{Max Planck Institue for Dynamics and Self-Organisation, G\"ottingen}
\affil[4]{Bernstein Center for Computational Neuroscience, G\"ottingen University}
\affil[5]{Department of Physics, Georg-August University, G\"ottingen}

\date{\today}

\begin{document}

\twocolumn[
        \begin{@twocolumnfalse}
                \maketitle
                \begin{abstract}
                    The critical state is assumed to be optimal for any computation in recurrent neural networks, because criticality maximizes a number of abstract computational properties.
                    We challenge this assumption by evaluating the performance of a spiking recurrent neural network on a set of tasks of varying complexity at - and away from critical network dynamics.
                    To that end, we developed a spiking network with synaptic plasticity on a neuromorphic chip.
                    We show that the distance to criticality can be easily adapted by changing the input strength, and then demonstrate a clear relation between criticality, task-performance and information-theoretic fingerprint.
                    Whereas the information-theoretic measures all show that network capacity is maximal at criticality, this is not the case for performance on specific tasks: Only the complex, memory-intensive tasks profits from criticality, whereas the simple tasks suffer from it. 
                    Thereby, we challenge the general assumption that criticality would be beneficial for any task, and provide instead an understanding of how the collective network state should be tuned to task requirement to achieve optimal performance. 
                \end{abstract}
                \hspace{2.0cm}
        \end{@twocolumnfalse}
]

\section{\label{sec:introduction}Introduction}

A central challenge in the design of an artificial network is to initialize it such that it quickly reaches optimal performance for a given task.
For recurrent networks, the concept of criticality presents such a guiding design principle~\cite{boedecker2012,bertschinger2004,legenstein2007,kinouchi2006,shew2013,delpapa2017,langton1990} (for feed-forward networks see e.g.~\cite{yam2000,thimm1995,goodfellow2016}).
At a critical point, typically realized as a second order phase transition between order and chaos or stability and instability, a number of basic processing properties are maximized, including sensitivity, dynamic range, correlation length, information transfer, and susceptibility~\cite{harris2002,munoz2017,wilting2018c,barnett2013,tkavcik2015}. 
Because all these basic properties are maximized, it is widely believed that criticality is optimal for task performance ~\cite{delpapa2017,bertschinger2004,boedecker2012,kinouchi2006,shew2013,munoz2017,munoz2018,langton1990}.

Tuning a system precisely to a critical point can be challenging.
Thus ideally, the system self-organizes to a criticality autonomously via local learning rules.
This is indeed feasible in various manners by modifying the synaptic strength depending on the pre- and postsynaptic neurons' activity only~\cite{levina2007,meisel2009,stepp2015,andrade2015,delpapa2017,tetzlaff2010,munoz2017,zierenberg2018,poil2012,shin2006}.
The locality of the learning rules is key for biological and artificial networks where global information (e.g. task performance error or activity of distant neurons) may be unavailable or costly to distribute.
Recently, it has been shown that specific local learning rules can even be harnessed more flexibly: 
A theoretical study suggests that recurrent networks with local, homeostatic learning rules can be tuned towards and away from criticality by simply adjusting the input strength~\cite{zierenberg2018}. This would enable one to sweep the entire range of collective dynamics from subcritical to critical to bursty, and assess the respective task performance.

Complementary to tuning collective network properties like the distance to criticality, local learning also enables networks to \emph{learn} specific patterns or sequences~\cite{hebb1949,hopfield1982,bi1998,markram1997}.
For example, \gls{stdp} shapes the connectivity, depending only on the timing of the activity of the pre- and postsynaptic neuron.
\Gls{stdp} is central for any sequence learning -- a central ingredient in language and motor learning~\cite{markram1997,bi1998}.
Such learning could strongly speed up convergence, and enables a preshaping of the artificial network - akin to the shaping of biological networks during development by spontaneous activity~\cite{loidolt_in_prep}.

Given diverse learning rules and task requirements, it may be questioned whether criticality is always optimal for processing, or whether each task may profit from a different state, as hypothesized in~\citep{wilting2018c}.
One could speculate that e.g. the long correlation time at criticality on the one hand enables long memory retrieval, but on the other hand could be unfavorable if a task requires only little memory. However, the precise relation between the collective state, and specific task requirements is unknown.

When testing networks, the observed network performance is expected to depend crucially on the choice of the task. How can one then characterize performance independently of a specific task, like e.g. classification or sequence memory?
A natural framework to characterize and quantify processing of any local circuit in a task-independent manner builds on information theory~\cite{wibral2015}:
Classical information theory enables us to \textit{quantify} the transfer of information between neurons, the information about the past input, as well as the storage of information~\cite{shannon1948,wibral2015,cover2012}.
The storage of information can be measured within the network or as read out from one neuron.
In addition, most recently classical mutual information is being generalized to more than two variables within the framework of \gls{pid}~\cite{williams2010,bertschinger2013,wibral2015,lizier2018}.
\Gls{pid} quantifies the unique and redundant contribution of each source variable to a target, but most importantly also enables a rigorous quantification of synergistic computation, a key contributor for any information integration~\cite{wibral2015,wibral2017,wibral2017b,bertschinger2013,williams2010}.
Thereby information theory is a key stepping stone when linking local computation within a network, with global task performance. 

Simulations of recurrent networks with plasticity become very slow with increasing size, because every membrane voltage and every synaptic strength has to be updated. 
To achieve an efficient implementation, physical emulation of synapses and neurons in electrical circuitry are very promising~\cite{mead1990,douglas1995}.
In such ``neuromorphic chips``, all neurons operate in parallel, and thus the speed of computation is largely independent of the system size, and is instead determined by the time constants of the underlying physical neuron and synapse models -- like in the brain.
Realizing such an implementation technically remains challenging, especially when using spiking neurons and flexible synaptic plasticity.
The BrainScaleS 2 prototype system combines physical models of neurons and synapses~\cite{aamir2018} with a general purpose processor carrying out plasticity~\cite{friedmann2017}.
In this system, the analog elements provide a speedup, energy efficiency, and enables scaling to very large systems, whereas the general purpose processor enables to set the desired learning rules flexibly.
Thus with this neuromorphic chip, we can run the long-term learning experiments --  required to study the network self-organization  -- within very short compute-time.
 
In the following, we investigate the relation between criticality, task-performance and information theoretic fingerprint.
To that end, we show that a spiking neuromorphic network with synaptic plasticity can be tuned towards and away from criticality by adjusting the input strength.
We show that criticality is beneficial for solving \emph{complex} tasks, but not the simple ones -- challenging the common notion that criticality in general is optimal for computation.
Methods from classical information theory as well as the novel framework of \glsfirst{pid} show that our networks indeed enfold their maximum capacity in the vicinity of the critical point.
Moreover, the lagged mutual information between the stimulus and the activity of neurons allows to establish a relation between criticality (as set by the input strength) and task-performance.
Thereby, we provide an understanding how basic computational properties shape task performance.

\section{\label{sec:results}Results}

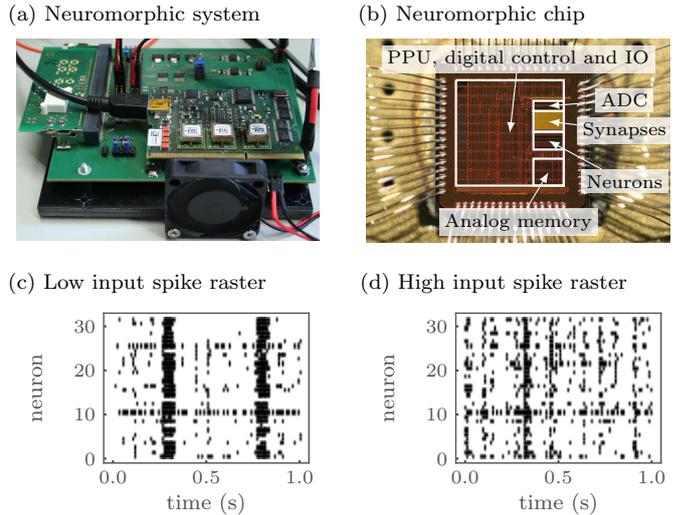
\begin{figure}[t]
	\subfloat[Neuromorphic system]{
		\centering
			\includegraphics[width=.45\columnwidth]{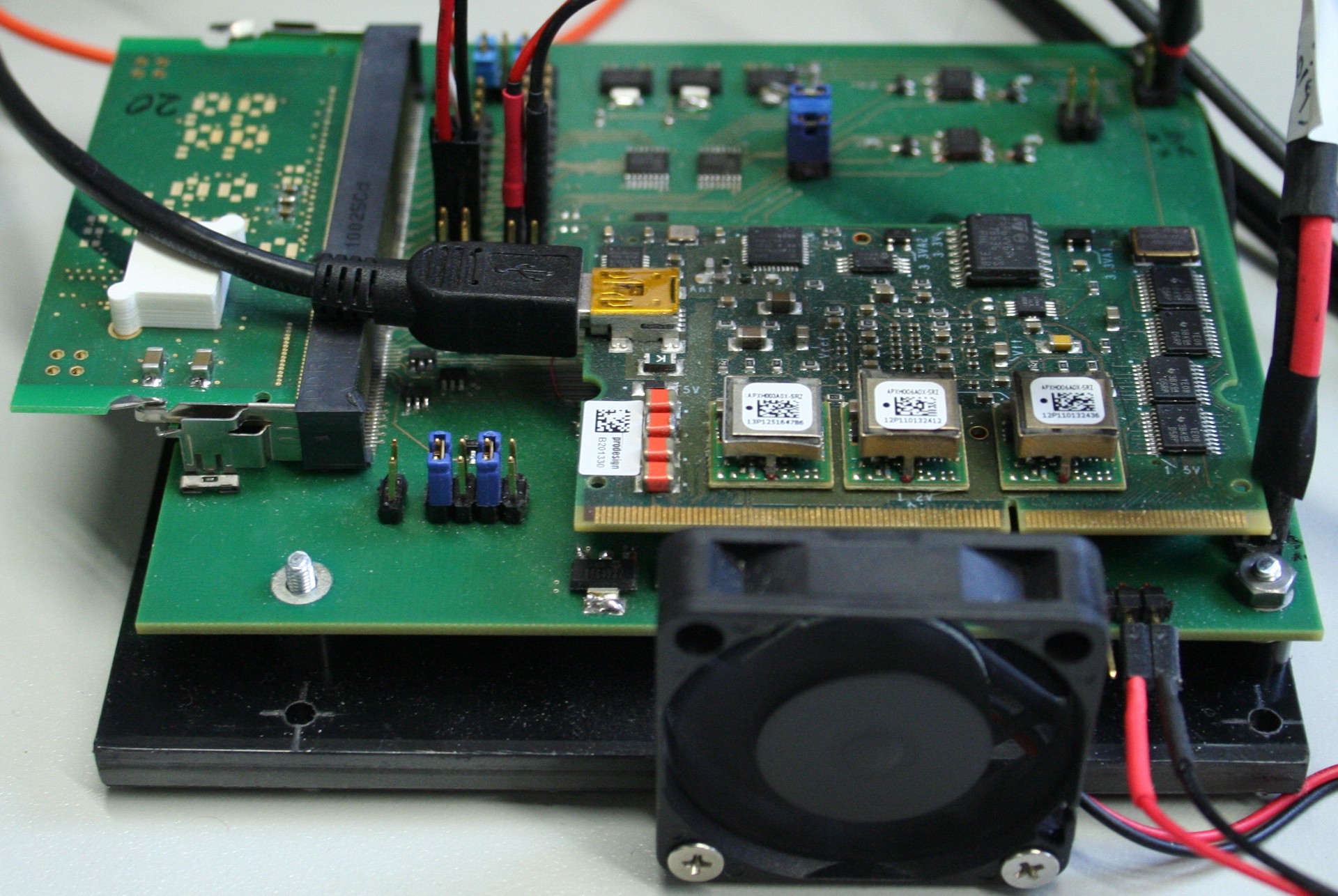}
			\label{fig:baseboard}
	}
	\hfill
    \subfloat[Neuromorphic chip]{
        \centering
            \begin{tikzpicture}[
            		label/.style = {fill=white, fill opacity=0.75, text opacity=1.0, inner sep=2pt}
            ]
            	\node[anchor=south west, inner sep=0] (image) at (0, 0) {
            		\includegraphics[width=0.45\columnwidth]{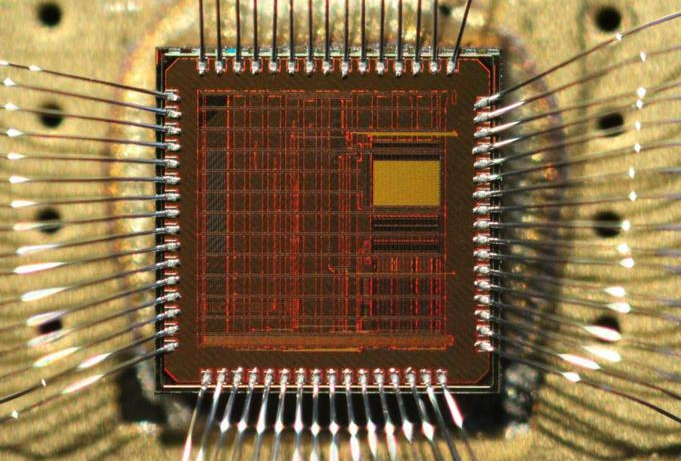}
            	};
            	\begin{scope}[x={(image.south east)},y={(image.north west)}]
            		\newcommand\ax{ 590.0/2000.0}
            		\newcommand\bx{1100.0/2000.0}
            		\newcommand\cx{1300.0/2000.0}
            		\newcommand\lx{\cx + 0.2}
            		\newcommand\ay{1420.0/1784.0}
            		\newcommand\by{1240.0/1784.0}
            		\newcommand\cy{1150.0/1784.0}
            		\newcommand\dy{ 970.0/1784.0}
            		\newcommand\ey{ 815.0/1784.0}
            		\newcommand\fy{ 735.0/1784.0}
            		\newcommand\gy{ 500.0/1784.0}
            
            		\draw[thick, color=white] (\ax,\ay) -- (\cx,\ay) -- (\cx,\gy) -- (\ax,\gy) -- (\ax,\ay); 
            		\draw[thick, color=white] (\bx,\by) rectangle (\cx,\cy); 
            		\draw[thick, color=white] (\bx,\cy) rectangle (\cx,\dy); 
            		\draw[thick, color=white] (\bx,\dy) rectangle (\cx,\ey); 
            		\draw[thick, color=white] (\bx,\fy) rectangle (\cx,\gy); 
            
            		\node[label] (ppu) at (0.5,0.9) {\scriptsize PPU, digital control and IO};
            		\node[label] (adc) at (\lx,0.70) {\scriptsize ADC};
            		\node[label] (syn) at (\lx,0.55) {\scriptsize Synapses};
            		\node[label] (nrn) at (\lx,0.30) {\scriptsize Neurons};
            		\node[label] (mem) at (0.5,0.10) {\scriptsize Analog memory};
            
            		\draw[-latex, color=white] (ppu) -- (\ax/2. + \cx/2, \ay/2 + \gy/2);
            		\draw[-latex, color=white] (adc) -- (\bx/2 + \cx/2, \by/2 + \cy/2);
            		\draw[-latex, color=white] (syn) -- (\bx/2 + \cx/2, \cy/2 + \dy/2);
            		\draw[-latex, color=white] (nrn) -- (\bx/2 + \cx/2, \dy/2 + \ey/2);
            		\draw[-latex, color=white] (mem) -- (\bx/2 + \cx/2, \fy/2 + \gy/2);
            	\end{scope}
            \end{tikzpicture}
            \label{fig:chip}
    }
	
	\subfloat[Low input spike raster]{
		\centering
			\includegraphics[width=.45\columnwidth]{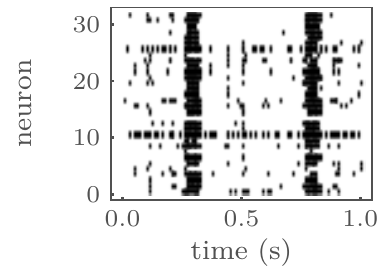}
			\label{fig:raster_low}
	}
	\hfill
	\subfloat[High input spike raster]{
		\centering
			\includegraphics[width=.45\columnwidth]{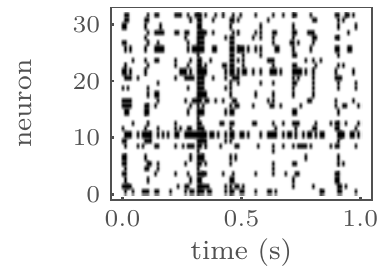}
			\label{fig:raster_high}
	}
	
	\caption{
 		The degree of external input $K_\mathrm{ext}$ shapes the collective dynamics of the network.
 		(a) The neural network is implemented on the prototype neuromorphic hardware system BrainScaleS 2.
 		(b) This system features an analog-neural network core as well as an on-chip general purpose processor that allows for flexible plasticity implementation.
 		(b) For low degree of input ($K_\mathrm{ext} = 0.25$), strong recurrent connections develop, and the  activity shows irregular bursts, resembling a critical state.
 		(c) For high degrees of input ($K_\mathrm{ext} = 0.56$), firing becomes more irregular and asynchronous.
	}
	\label{fig:raster}
\end{figure}

\textbf{Model overview.} We emulate networks of \gls{lif} neurons on the mixed-signal neuromorphic prototype system\footnote[1]{
    Future versions of the BrainScaleS 2 chip will feature 512 neuron circuits with adaptive-exponential \acrshort{lif} dynamics and inter-compartmental conductances.
}
 BrainScaleS 2, which has $N=32$ neurons (\cref{fig:baseboard,fig:chip,tab:parameters}).
We use the term \textit{emulation} in order to clearly distinguish between the physical implementation, where each observable has a measurable counterpart on the neuromorphic chip, and standard software \textit{simulations} on conventional hardware.
The system features an array of $32\times 32$ current-based synapses, where $20\%$ of the synapses are programmed to be inhibitory.
Synaptic plasticity acts equally on all synapses and is composed of a positive drift and a negative anticausal \gls{stdp} term.
In conjunction both terms lead to homeostatic regulation and thus stable network activity of about \SI{20}{\hertz} per neuron (see figure~1a of Supplemental Material~\cite{supplement}).
Plasticity is executed by an on-chip general purpose processor alongside to the analog emulation of neurons and synapses.
This allows for an uninterrupted and fast data acquisition.
Even for the small prototype system, the advantages of neuromorphic computing in terms of speed and energy efficiency become important as depicted in~\cite{wunderlich2019}.

Neurons are potentially all-to-all connected, but $K_\mathrm{ext}$ out of the $N$ synapses per neuron are used to inject external Poisson or pattern input.
Effectively, $K_\mathrm{ext}$ quantifies the input strength with the extreme cases of $K_\mathrm{ext} / N = 1$ for a feed-forward network and $K_\mathrm{ext} / N = 0$ for a fully connected recurrent network, which is completely decoupled from the input.
Depending on the degree of external input $K_\mathrm{ext}$, the network shows diverse dynamics (\cref{fig:raster_low,fig:raster_high}).
As expected~\cite{zierenberg2018}, $K_\mathrm{ext}$ shapes the collective dynamics of the network from synchronized for low $K_\mathrm{ext}$ to more asynchronous-irregular for high $K_\mathrm{ext}$.

\begin{figure}[t]
	\subfloat[Avalanche distribution]{
		\centering
			\includegraphics[width=.45\columnwidth]{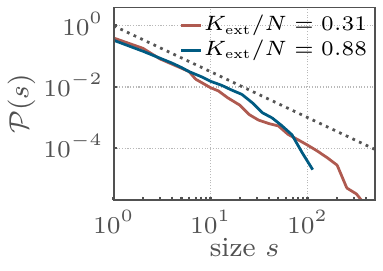}
			\label{fig:distribution}
	}
	\hfill
	\subfloat[Exponential cutoff]{
		\centering
			\includegraphics[width=.45\columnwidth]{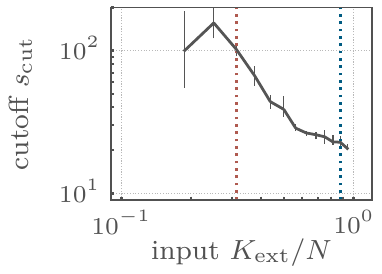}
			\label{fig:cutoff}
	}
	
	\subfloat[Critical exponents]{
		\centering
			\includegraphics[width=.45\columnwidth]{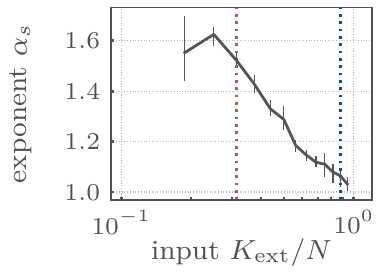}
			\label{fig:exponents}
	}
	\hfill
	\subfloat[Model comparison]{
		\centering
			\includegraphics[width=.45\columnwidth]{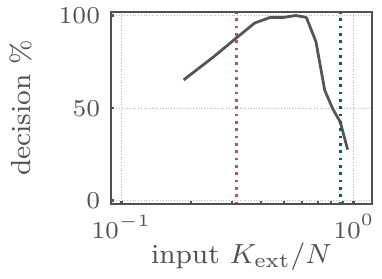}
			\label{fig:comparison}
	}
	
	\caption{%
	    Under low degree of input $K_\mathrm{ext}$, the network self-organizes towards a critical state, and shows long-tailed avalanche distributions.
	    (a) Distributions of avalanche sizes $s$ show power-laws over two orders of magnitude for low $K_\mathrm{ext}$.
	    Fitting a truncated power law, (b) the exponential cutoff $s_\mathrm{cut}$  peaks, and (c) critical exponents $\alpha_s$  approximate $1.5$, as expected for critical branching processes.
	    (d) A maximum-likelihood comparison decides for a power-law compared to an exponential fit in the majority of cases. 
	    Dashed vertical lines indicate the set of $K_\mathrm{ext} / N$ values that have been selected in (a).
	    In this and all following figures, the median over runs and (if acquired) trials is shown, and the errorbars show the $5\%$-$95\%$ confidence intervals.
	}
	\label{fig:avalanche}
\end{figure}

\begin{figure}[t]
	\subfloat[Avalanche distribution]{
	    \centering
			\includegraphics[width=.45\columnwidth]{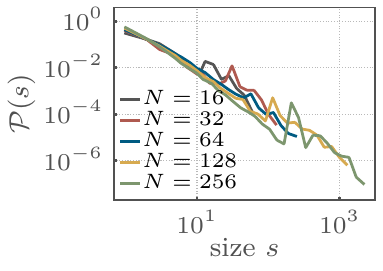}
			\label{fig:simulation_avalanches}
	}
	\hfill
	\subfloat[Finite-size scaling]{
	    \centering
			\includegraphics[width=.45\columnwidth]{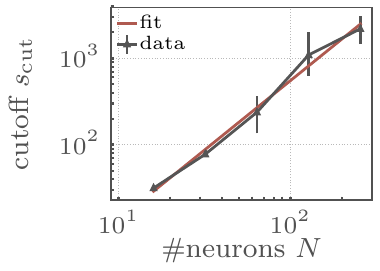}
			\label{fig:finite_size}
	}
	
	\caption{%
	    Finite-size scaling is assessed using a software implementation with varying system size $N$.
 		(a) Exemplary avalanche size distributions follow a power-law for any tested $N$ (degree of input $K_\mathrm{ext} / N = 1/4$).
 		(b) As expected for critical systems, the cutoff $s_\mathrm{cut}$ scales with the system size. The scaling exponent is \SI{1.6(2)}{}.
	}
	\label{fig:software_simulations}
\end{figure}

\begin{figure}[t]
	\subfloat[Branching ratio]{
		\centering
			\includegraphics[width=.45\columnwidth]{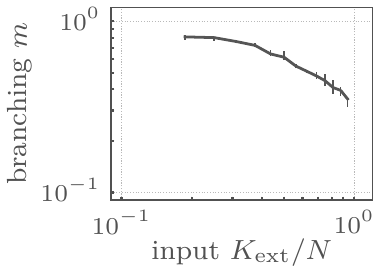}
			\label{fig:branch}
	}
	\hfill
	\subfloat[Autocorrelation time]{
		\centering
			\includegraphics[width=.45\columnwidth]{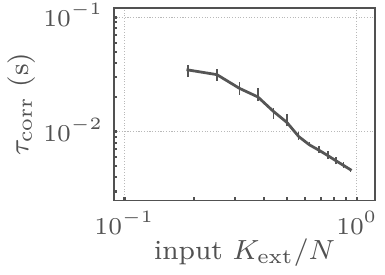}
			\label{fig:corr}
	}
	
	\subfloat[Model validation]{
		\centering
			\includegraphics[width=.45\columnwidth]{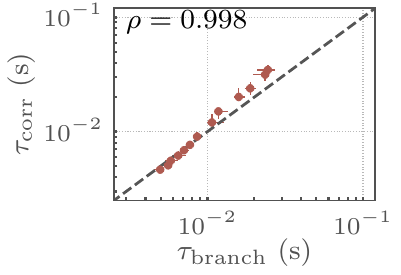}
			\label{fig:corrcorr}
	}
	\hfill
	\subfloat[Variations]{
		\centering
			\includegraphics[width=.45\columnwidth]{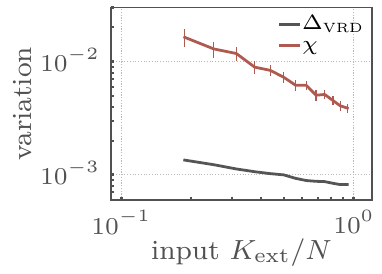}
			\label{fig:variation}
	}
	
	\caption{%
	    For low degree of input $K_\mathrm{ext}$, the network shows clear signatures of criticality beyond power-laws.
	    Only for low values of $K_\mathrm{ext}$, (a) the estimated branching ratio $m$ tends towards unity, and (b) the estimated autocorrelation time  $\tau_\mathrm{corr}$ peaks.
	    (c) The match of the  $\tau_\mathrm{corr}$, and the $\tau_\mathrm{branch} \sim -1 / \log{(m)}$ as inferred from $m$ supports the criticality hypothesis (correlation coefficient of $\rho = 0.998$, $p < 10^{-10}$).
	    (d) Trial-to-trial $\Delta_\mathrm{VRD}$ variations as well as the susceptibility $\chi$ increase for low $K_\mathrm{ext}$.
	}
	\label{fig:classical}
\end{figure}

\textbf{Critical dynamics arise under low input $K_\mathrm{ext}$.}
The transition to burstiness for low $K_\mathrm{ext}$ suggests the emergence of critical dynamics, i.e.  dynamics expected at a non-equilibrium second order phase transition.
Indeed, as detailed in the following, we find signatures of criticality in the classical avalanche distributions (\cref{fig:avalanche,fig:software_simulations}) as well as in the branching ratio (\cref{fig:branch}), the autocorrelation time (\cref{fig:corr}), the susceptibility and in trial-to-trial variations (\cref{fig:variation}).

To test whether the network indeed approaches criticality, we assume the established framework of a branching process~\cite{zapperi1995,harris2002,watson1875,wilting2018a}.
In branching processes, a spike at time $t$ triggers on average $m$ postsynaptic spikes at time $t + 1$, where $m$ is called the branching parameter.
For $m = 1$ the process is critical, and the dynamics give rise to large cascades of activity, called \textit{avalanches}~\cite{beggs2003,harris2002}.
The size $s$ of an avalanche is the total number of spikes in a cluster and is power-law distributed at criticality.
The binwidth for the estimation of the underlying distributions is set to the mean inter-event interval following common methods~\cite{priesemann2018can}.
Our network shows power-law distributed avalanche sizes $s$ over two orders of magnitude for low $K_\mathrm{ext}$ (\cref{fig:distribution}).
For almost any  $K_\mathrm{ext}$, the distribution is better fitted by a power-law than by an exponential distribution~\cite{clauset2009} (\cref{fig:comparison}). However, only for low $K_\mathrm{ext}$ the exponent of the avalanche distribution is close to the expected one, $\alpha_s \approx 1.5$ (\cref{fig:exponents}), and the power-law shows the largest cutoff $s_\mathrm{cut}$ (\cref{fig:cutoff}).
For low $K_\mathrm{ext}$, the networks tend to get unstable due to the limited number of neurons explaining the decline in $s_\mathrm{cut}$ (\cref{fig:cutoff}) and in the maximum likelihood comparison (\cref{fig:comparison}).
Together, all the quantitative assessment of the avalanches indicate that a low degree of input $K_\mathrm{ext}$ produces critical-like behavior.

In a control experiment, we investigate finite-size scaling in software simulations, as the current physical system features only 32 neurons.
Therefore, a network with the same topology, plasticity rules and single neuron dynamics (though without parameter noise and hardware constraints) is simulated for various system sizes $N$.
The resulting avalanche distributions show power-laws for any system size (\cref{fig:simulation_avalanches}), and the cutoff $s_\mathrm{cut}$  scales with $N$ as expected at criticality (\cref{fig:finite_size}).
The  scaling exponent is \SI{1.6(2)}{}.
Together, these numerical results confirm the hypothesis that for low degrees of input $K_\mathrm{ext}$, the small network that is emulated on the chip self-organizes as close to a critical state as possible.

The implementation on neuromorphic hardware promises fast emulation.
Already for $N=32$, the neuromorphic chip is about a factor of 100 faster than the Brian 2 simulation.
To give numbers, a single plasticity experiment with a duration of \SI{600}{\second} biological time is simulated in \SI{570}{\second} in Brian 2, but emulated in only \SI{6}{\second} on the neuromorphic chip.
Hence, a neuromorphic implementation is very promising especially for the future full size chip:
When running such detailed networks as classical simulations, the computational overhead scales with $\mathcal{O}(N^2)$ due to the all-to-all connectivity and synaptic plasticity on conventional hardware.
In contrast, for the neuromorphic system, the execution time is largely independent of the system size $N$, as long as the network can be implemented on the system.

The assumption that the critical state of the network corresponds to the universality class of critical branching processes is tested further by inferring the branching parameter $m$ (\cref{eq:ar}) proper, the autocorrelation times and the response to perturbations.
First, the branching parameter $m$ characterizes the spread of activity and is smaller (larger) than unity for sub-critical (super-critical) processes.
For our model, it is always in the sub-critical regime, but tends towards unity for low $K_\mathrm{ext}$ (\cref{fig:branch}).
Second, the autocorrelation time $\tau_\mathrm{corr}$ is expected to diverge at criticality as $\tau_\mathrm{branch}(m) \sim \lim\limits_{m\rightarrow 1} (- 1 / \log{(m)}) = \infty$~\cite{wilting2018a}.
Indeed, $\tau_\mathrm{corr}$ as estimated directly from the autocorrelation of the population activity is maximal for low $K_\mathrm{ext}$ (\cref{fig:corr}).
Third, the estimates of $m$ and $\tau_\mathrm{corr}$ are in theory related via the analytical relation $\tau_\mathrm{branch}(m) \sim - 1/ \log{(m)}$. This relation holds very precisely in the model (\cref{fig:corrcorr}, correlation coefficient $\rho=0.998, p<10^{-10}$).
Fourth, towards criticality, the response to any perturbation increases. The impact of a small perturbation is quantified by a variant of the van-Rossum distance $\Delta_\mathrm{VRD}$ (\cref{eq:distance}). It peaks for low degrees of the external input $K_\mathrm{ext}$ (\cref{fig:variation}). Last, one advantage of operating in the vicinity of a critical point is the ability to enhance stimulus differences by the system response.
This is reflected in a divergence of the susceptibility at the critical point.
The susceptibility $\chi$ (\cref{eq:susceptibility}), quantified here as the change in the population firing rate in response to a burst of $N_\mathrm{pert} = 6$ additional spikes, is highest for low $K_\mathrm{ext}$ (\cref{fig:variation}).
Thus overall, the avalanche distributions as well as the dynamic properties of the network all indicate that it self-organizes to a critical point under low degree of input $K_\mathrm{ext}$. 

\begin{figure}[t]
	\subfloat[$n$-bit sum, $N_\mathrm{read} = 16$]{
		\centering
			\includegraphics[width=0.45\columnwidth]{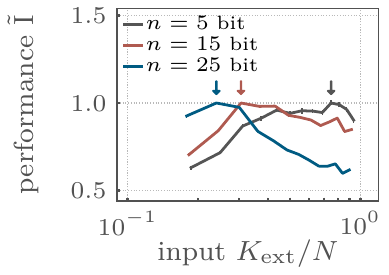}
			\label{fig:sum}
	}
	\hfill
	\subfloat[$n$-bit parity, $N_\mathrm{read} = 16$]{
		\centering
    		\includegraphics[width=0.45\columnwidth]{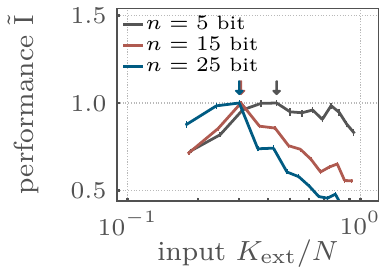}
			\label{fig:parity}
	}
	
	\subfloat[$N_\mathrm{read} = 8$]{
		\centering
    		\begin{tikzpicture}
    			\node[anchor=north west] (plot) at (0,0) {\includegraphics[width=0.44\columnwidth]{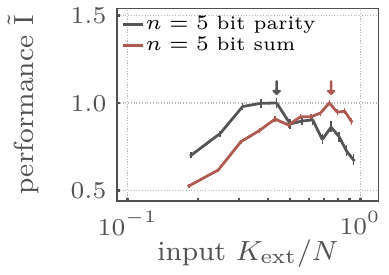}};
    			\draw[latex-, thick] (2.5, -2) -- (3.5, -2) node[midway,above] {\scriptsize Critical};
    		\end{tikzpicture}
			\label{fig:sub_simple}
	}
	\hfill
	\subfloat[$N_\mathrm{read} = 4$]{
		\centering
			\includegraphics[width=.45\columnwidth]{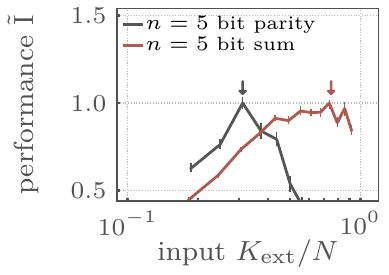}
			\label{fig:sub_complex}
	}
	
	\caption{%
		Computational challenging tasks profit from critical network dynamics (small $K_\mathrm{ext}$)-- simple tasks do not.
		The network is used to solve (a) a $n$-bit sum and (b) a $n$-bit parity task by training a linear classifier on the activity of $N_\mathrm{read} = 16$ neurons.
		Here, task complexity increases with $n$, the number of past inputs that need to be memorized and processed.
		For high $n$, task performance profits from criticality, whereas simple tasks suffer from criticality.
		Especially, the more complex, non-linear parity tasks profits from criticality.
		Task complexity can also be increased by further restricting the classifier to (c) $N_\mathrm{read} = 8$ and (d) $N_\mathrm{read} = 4$.
		Again, the parity task increasingly profits from criticality with decreasing $N_\mathrm{read}$.
		The performance is quantified by the normalised mutual information $\mathrm{\tilde{I}}$ between the vote of the classifier and the parity or sum of the input.
		Highest performance for a given task is highlighted by colored arrows.
	}
	\label{fig:task}
\end{figure}

\textbf{Network properties have to be tuned to task requirements for optimal performance.}
It is widely assumed that criticality optimizes task performance.
However, we found that one has to phrase this statement more carefully.
While certain abstract computational properties, like the susceptibility, sensitivity or memory time span are indeed maximal or even approaching infinity at a critical state, this is not necessary for task performance in general~\cite{shew2011,shew2013,barnett2013,wilting2018a}.
We find that it can even be detrimental.
For every single task complexity, a different distance to criticality is optimal, as outlined in the following.

We study the performance of our recurrent neural network in the framework of \textit{reservoir}:
The performance of a recurrent neural network is quantified by the ability of linear readout neurons to separate different sequences~\cite{maass2002,jaeger2001,schurmann2005}.
To that end, it is often necessary to maintain information about past input for long time spans.
To test performance, we specifically use a $n$-bit sum and a $n$-bit parity task and trained a readout on the activity of $N_\mathrm{read} = 16$ randomly chosen neurons of the reservoir.
For the two given tasks complexity increases with $n$: to solve the tasks, the network has to both \textit{memorize} and \textit{process} the input from the $n$ past steps.
As reservoirs close to a critical point have longer memory as quantified by the lagged mutual information ($I_\tau$, \cref{fig:corr}), one expects that particularly the memory intensive tasks profit from criticality (tasks with high $n$ are better at low degrees of input $K_\mathrm{ext}$).
In contrast, simple tasks (low $n$) might suffer from criticality because of the maintenance of memory about unnecessary input. 
Since the estimation of parity, in contrast to the sum, is fully non-linear, their direct comparison allows to further dissect task complexity.
Thus, depending on the task complexity, there should be an ideal $K_\mathrm{ext}$, leading to maximal performance.

For our network, we find indeed that maximal task performance depends on both, task complexity and distance to criticality: simple sum tasks ($n=5$) are optimally solved away from criticality, whereas complex sum tasks ($n=25$) profit from the long timescales arising at criticality (\cref{fig:sum}).
The non-linear parity task profits even more from criticality: even for $n=5$ networks closer to the critical point promote task performance (\cref{fig:parity}).
Hence, we are capable of adapting the networks computational properties to task complexity by fine-tuning the strength of the input.

To further tune the difficulty of task, we reduced the number neurons visible to the readout $N_\mathrm{read}$.
We expect that in principle information about e.g. parity could be available in a single neuron if the network is sufficiently close to criticality, because critical network dynamics are not only characterized by temporal, but also spatial correlations.
The ability to condense information about extended stimuli in the activity of few neurons can be valuable.
To quantify the effect of spatial correlations on computation, we trained linear classifiers on the activity of a subset of neurons for the $5$-bit sum and the parity task.
When lowering $N_\mathrm{read}$ from 8 to 4, only the non-linear parity tasks increasingly profits from critical network dynamics (\cref{fig:sub_simple,fig:sub_complex}).
In contrast, the information necessary to solve the linear sum task seems to be globally available in the network response even for sub-critical dynamics.
The ability to locally read out global information from the network is of equal importance for both, large neuromorphic systems~\cite{schemmel2010} and living networks~\cite{brette2019,bernardi2017}.

\begin{figure}[t]
	\subfloat[5-bit parity task]{
		\centering
			\includegraphics[width=0.45\columnwidth]{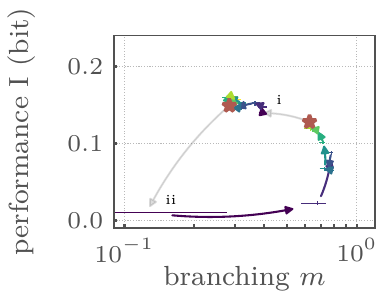}
			\label{fig:switch_simple}
	}
	\hfill
	\subfloat[15-bit parity task]{
		\centering
			\includegraphics[width=0.45\columnwidth]{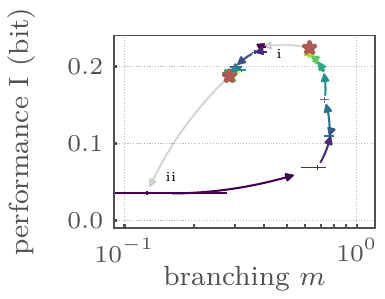}
			\label{fig:switch_complex}
	}
	
	\subfloat{
		\centering
			\includegraphics[width=.95\columnwidth]{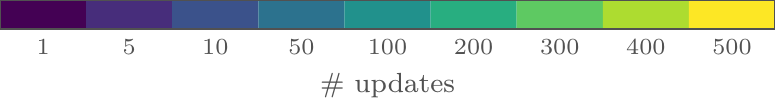}
	}

	\caption{%
		The network can be dynamically adapted by changing the degree of the input $K_\mathrm{ext}$.
		After convergence of synaptic weights $w_{ij}$, $K_\mathrm{ext}/N$ is (\romannum{1}) switched from critical ($K_\mathrm{ext} = \SI{0.3}{}$) to sub-critical ($K_\mathrm{ext} = \SI{0.8}{}$) and (\romannum{2}) vice versa.
		The branching ratio $m$ and the performance $\mathrm{I}$ of the network on (a) the $5$-bit and (b) the $15$-bit parity task are evaluated after various numbers of synaptic updates.
		The network reaches the same performance and dynamics as when starting from $w_{ij} = 0\,\forall\,i,j$ (marked by red stars).
		For both tasks, the transition from sub-critical to critical dynamics requires more updates as expected.
		Moreover, optimal performance for (a) the $5$-bit task is achieved under strong input (\romannum{1}), whereas for (b) the $15$-bit task requires low input (\romannum{2}).
		The performance is quantified by the mutual information $\mathrm{I}$ between the parity of the input and the vote of a linear classifier.
	}
	\label{fig:switch}
\end{figure}

\textbf{The adaptation to task can be achieved by dynamically switching the input strength.}
We know from the previous experiments that for high $n$, the $n$-bit parity task is solved best at criticality, whereas for low $n$, the sub-critical regime leads to best performance.
In the following, we investigate how to transit between both states.
To achieve this, we take the state of a critical network and switch the degree of the input $K_\mathrm{ext}$ to a sub-critical configuration and vice versa.
The performance is evaluated after various numbers of synaptic updates.
This task switch generates the same working points as the previous emulations that start with synaptic weights $w_{ij} = 0\,\forall\,i, j$ and have a long adaptation phase (red stars in \cref{fig:switch}).

A fast adaption to different input strengths is required to switch between tasks of different complexity.
The transition from critical to sub-critical is achieved after the application of about 50 synaptic updates corresponding to \SI{50}{\second} biological time, whereas going from sub-critical to critical takes about 500 updates and therefore \SI{500}{\second} (\cref{fig:switch}).
However, due to the speedup of the neuromorphic chip, the adaptation takes only about \SI{0.5}{\second} wall clock time and can even be lowered by decreasing the integration time over spike-pairs in the synaptic update rule.
As alternative strategies, one could switch between saved configurations, or run a hierarchy of networks with different working points in parallel~\cite{zierenberg2019}.

\begin{figure}[t]
	\subfloat[Lagged input-neuron I]{
		\centering
			\includegraphics[width=.45\columnwidth]{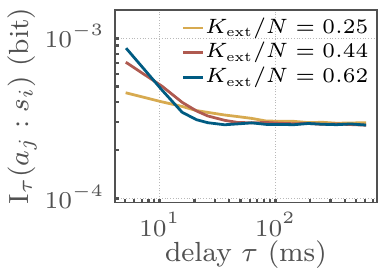}
			\label{fig:laggedMI_external}
	}
	\hfill
	\subfloat[Input-neuron MC]{
		\centering
			\includegraphics[width=.45\columnwidth]{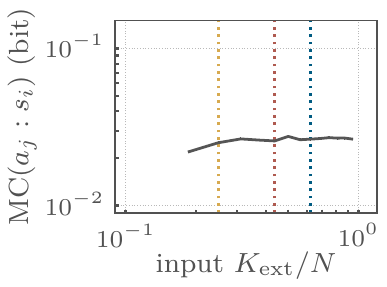}
			\label{fig:mc_external}
	}
	
	\subfloat[Lagged neuron-neuron I]{
		\centering
			\includegraphics[width=.45\columnwidth]{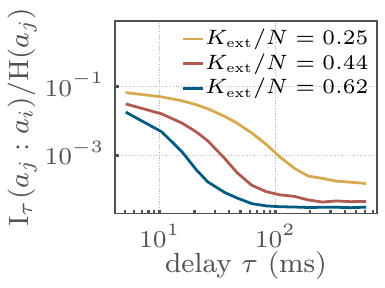}
			\label{fig:laggedMI_internal}
	}
	\hfill
	\subfloat[Neuron-neuron MC]{
		\centering
			\includegraphics[width=.45\columnwidth]{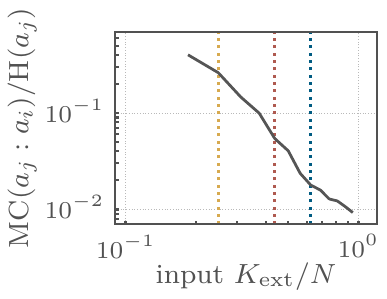}
			\label{fig:mc_internal}
	}
	
	\caption{%
	    	Long lasting memory accompanies critical network dynamics.
		(a) Memory about the input $s_i$ as read out from neuron $a_j$ after a time lag $\tau$ is quantified by the mutual information $\mathrm{I}_\tau(a_j, s_i)$.
		Here, high degrees of the external input $K_\mathrm{ext}$ are favorable for memory on short timescales, whereas low $K_\mathrm{ext}$ is favorable on larger timescales. 
		(b) The \glsfirst{mc} stays  fairly constant, despite of a decreased coupling to the stimulus for low $K_\mathrm{ext}$.
		(c) The lagged I between the activity of pairs of neurons indicates increasing memory for decreasing $K_\mathrm{ext}$, also visible in the \gls{mc} (d).
		The selection of $K_\mathrm{ext} / N$ in (a) and (c) is marked by dashed vertical lines in (b) and (d).
	}
	\label{fig:memory_capacity}
\end{figure}

\textbf{Information theory enables task-independent quantification of computational properties.} While task performance is the standard bench-mark for any model, such bench-mark tasks have two disadvantages: In many biological systems, like higher brain areas or in \textit{vitro} preparations, such tasks cannot be applied.
Even if tasks can be applied, the outcome will always depend on the chosen task. 
To quantify computational properties in a \textit{task-independent} manner, information theory offers powerful tools~\cite{wibral2015}.
Using the Poisson noise input, we find that the lagged mutual information $\mathrm{I}_\tau$ between the input $s_i$ and the activity of a neuron after a time lag $\tau$, $a_j$ predicts the performance on the parity task.
Here, at high $K_\mathrm{ext}$ (away from criticality) information about the input is maximal for very short $\tau$, but decays very quickly (\cref{fig:laggedMI_external}).
This \textit{fast-forgetting} is important to irradiate past, task-irrelevant input that would interfere with novel, task-relevant input.
At small $K_\mathrm{ext}$, the recurrence is stronger and input can be read out for much longer delays (\SI{20}{\milli\second} vs. \SI{60}{\milli\second}).
This active storage of information is required in a reservoir to solve any task that combines past and present input, and hence the more complex parity task also profits from it.
However, the representation of input in every single neuron becomes less reliable (i.e. $I_\tau$ is smaller).
A measure for the representation of the input in the network could be obtained by integrating $\mathrm{I}_\tau$ over $\tau$. 
Interestingly, this \glsfirst{mc} stays fairly constant  (\cref{fig:mc_external}).
Note that we only quantified the representation of the input in a \textit{single neuron}, a measure very easily accessible in experiments; obviously the readout can draw on the distributed memory across all neurons, which jointly provide a much better readout.

The memory maintenance for task processing has to be realized mainly by activity propagating on the recurrent connections in the network. Therefore, it is often termed \gls{ais}~\cite{Lizier2008,wibral2014}.
The recurrent connections become stronger closer to criticality, and as a consequence we find that the lagged mutual information between pairs of neurons in the \textit{reservoir} also increases (\cref{fig:laggedMI_internal}).
As a result the \gls{mc} of the reservoir increases over almost two orders of magnitude when approximating criticality (lower $K_\mathrm{ext}$, \cref{fig:mc_internal}).
This increase in \textit{internal} \gls{mc} carries the performance on the more complex parity tasks. 

\begin{figure}[t]
	\subfloat[Entropy]{
		\centering
			\includegraphics[width=.45\columnwidth]{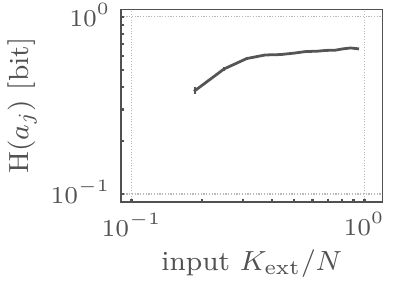}
			\label{fig:entropy}
	}
	\hfill
	\subfloat[Mutual information]{
		\centering
			\includegraphics[width=.45\columnwidth]{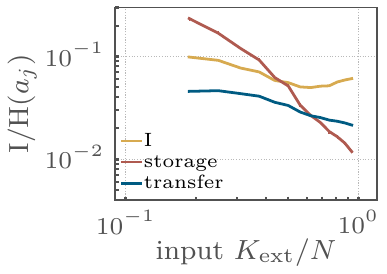}
			\label{fig:mi}
	}
	
	\caption{%
 		The information fingerprint changes with the degree of input $K_\mathrm{ext}$, thus with distance to criticality.
 		(a) The \glsfirst{h} of the spiking activity of a single neuron, $a_j$ stays fairly constant, except for low $K_\mathrm{ext}$ as a consequence of decreasing firing rates.
 		(b) The \glsfirst{i} between the activity of two units $a_i$, $a_j$ increases  with lower $K_\mathrm{ext}$ (i.e. closer to critical).
 		The network intrinsic memory also increases, indicated by the \glsfirst{ais} $\mathrm{I}(a_j:\mathbf{a_j^-})$.
 		Likewise, the information transfer within the network increases with lower $K_\mathrm{ext}$.
 		Information transfer is measured as \glsfirst{te} between pairs of neurons $a_j$ and $a_i$, $\mathrm{I}(a_j: \mathbf{a}_i^-\vert \mathbf{a}_j^-)$.
	}
	\label{fig:info}
\end{figure}

\begin{figure}[t]
	\subfloat[PID components]{
    	\centering
    	\hspace{0.25cm}
        \begin{tikzpicture}[scale=0.8]
        	\def\syn{(2.0,2.0) ellipse (85pt and 35pt)}
        	\def\unqeins{(1.0,2.0) ellipse (40pt and 20pt)}
        	\def\unqzwei{(3.0,2.0) ellipse (40pt and 20pt)}
        	\begin{scope}
        		\clip\syn;
        		\draw[fill=nicegray]\syn;
        	\end{scope}
        	\begin{scope}
        		\clip\unqeins;
        		\draw[fill=niceyellow]\unqeins;
        	\end{scope}
        	\begin{scope}
        		\clip\unqzwei;
        		\draw[fill=niceblue]\unqzwei;
        	\end{scope}
        	\begin{scope}
        		\clip\unqeins;
        		\draw[fill=nicered]\unqzwei;
        	\end{scope}
        	\draw\syn;
        	\draw\unqeins;
        	\draw\unqzwei;
        	\node (N1) at ( 0.40, 4.50) {\footnotesize$\mathrm{I}_\mathrm{shd}(a_j: \mathbf{a_j^-}; \mathbf{a_i^-})$};
        	\draw[-latex, thick] (N1) -- (2.0, 2.0);
        	\node (N2) at (-1.0, 3.50) {\footnotesize$\mathrm{I}_\mathrm{unq}(a_j: \mathbf{a_j^-} \setminus \mathbf{a_i^-})$};
        	\draw[-latex, thick] (N2) -- (0.5, 2.0);
        	\node (N3) at (5.0, 3.50) {\footnotesize$\mathrm{I}_\mathrm{unq}(a_j: \mathbf{a_i^-} \setminus \mathbf{a_j^-})$};
        	\draw[-latex, thick] (N3) -- (3.5, 2.0);
        	\node (N4) at ( 2.00, -0.15) {\footnotesize$\mathrm{I}(a_j: \mathbf{a_j^-}, \mathbf{a_i^-})$};
        	\draw[-latex, thick] (N4) -- (2.0, 0.78);
        	\node (N5) at (-0.25, -0.15) {\footnotesize$\mathrm{I}(a_j: \mathbf{a_j^-})$}; 
        	\draw[-latex, thick] (N5) -- (0.5, 1.35);
        	\node (N6) at ( 4.25, -0.15) {\footnotesize$\mathrm{I}(a_j: \mathbf{a_i^-})$}; 
        	\draw[-latex, thick] (N6) -- (3.5, 1.35);
        	\node (N7) at (3.5, 4.50) {\footnotesize$\mathrm{I}_\mathrm{syn}(a_j: \mathbf{a_j^-}, \mathbf{a_i^-})$};
        	\draw[-latex, thick] (N7) -- (2.1, 2.75);
        \end{tikzpicture}
    	\label{fig:pid_schematic}
    }
    
	\subfloat[Joint mutual information]{
		\centering
			\includegraphics[width=.45\columnwidth]{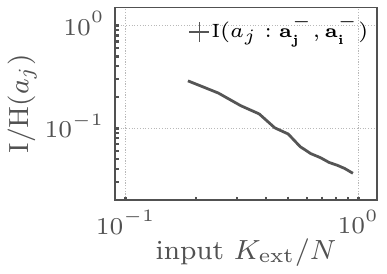}
			\label{fig:jointMI}
	}
    \hfill
	\subfloat[PID components]{
		\centering
			\includegraphics[width=.45\columnwidth]{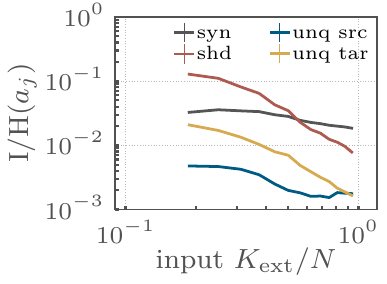}
			\label{fig:pid_internal}
	}
	\caption{%
	    \Glsfirst{pid} components increase towards criticality (i.e. with smaller input $K_\mathrm{ext}$). 
		(a) The two input variables for \gls{pid} correspond to the spiking histories $\mathbf{a_i^-}$ and $\mathbf{a_j^-}$ of two neurons, and the output variable to the present state $a_j$. \Gls{pid} enables to quantify the unique contribution of each source to the firing of the target neuron, as well as the shared (also called redundant), and synergistic contributions.
	    (b) The joint \glsfirst{i} increases with decreasing $K_\mathrm{ext}$.
	    (c) All \gls{pid} components increase with approaching criticality. Interestingly, the synergistic and shared contributions are always much larger than the unique contributions (note the logarithmic axis). This highlights the collective nature of processing in recurrent neural networks.
	}
	\label{fig:pidinternal}
\end{figure}

When assessing computational capacities, information theory enables us to quantify not only the \gls{h} and \gls{i} between units, but also to disentangle transfer and storage of information, as well as unique, redundant and synergistic contributions of different source neurons~\cite{wibral2015,wibral2017,schreiber2000,williams2010,bertschinger2013}.
We find that all these quantities increase with approaching criticality (smaller $K_\mathrm{ext}$, \cref{fig:mi,fig:pid_internal}).
This indicates that the overall computational capacity of the model increases, as predicted for the vicinity of the critical state~\cite{boedecker2012,barnett2013,bertschinger2004,langton1990}. 

In more detail, the \gls{ais} of a neuron, as well as the \gls{i} and the \gls{te} between pairs of neurons increase with lower $K_\mathrm{ext}$ (\cref{fig:mi}).
In our analysis, these increases reflect memory that is realized as activity propagation on the network, and not storage within a single neuron, because the bin-size used for analysis is larger than the refractory period $\tau_\mathrm{ref}$, synaptic- $\tau_\mathrm{syn}$ and membrane-timescales $\tau_m$. 
Information theory here enables us to show that active transfer and storage of information within the network strongly increases towards criticality.
A similar increase in \gls{i}, \gls{ais} and \gls{te} has been observed for the Ising model and reservoirs at criticality~\cite{barnett2013,boedecker2012}, and hence supports the notion that criticality maximizes information processing capacity.
Note however, that this maximal capacity is typically not necessary; as shown here, it can even be unfavorable when solving simple tasks.

Very recently, it has become possible to dissect further the contributions of different neurons to processing, using \gls{pid}~\cite{williams2010}.
\Gls{pid} enables us to disentangle for a target neuron  $a_i$, how much unique information it obtains from its own past activity $\mathbf{a}_i^-$, or the past activity of a second neuron $\mathbf{a}_j^-$; and how much information is redundant or even synergistic from the two (\cref{fig:pid_schematic}). Synergistic information is that part of information that can only be computed if \textit{both} input variables are known, whereas redundant information can be obtained from one or the other.

All the \gls{pid} components increase when approaching the critical point (low $K_\mathrm{ext}$, \cref{fig:pid_internal}).
Quantitatively, the redundant and the synergistic information are always stronger than the unique ones which are about ten times less. 
The shared information dominates closer to criticality, mirroring the increased network synchrony and redundancy between neurons.
Further, the synergistic contribution, i.e. the contributions that rely on the past of both neurons slightly increases, and is indeed the largest contribution for high $K_\mathrm{ext}$.
This reflects that typically the joint activity of both neurons is required to activate a \gls{lif} neuron.
Interestingly, the strong increase in shared information (i.e. redundancy) does not seem to impede the performance at criticality (small $K_\mathrm{ext}$).
However, for even higher synchrony, as expected beyond this critical transition, the shared information might increase too much and thereby decrease performance.

\section{\label{sec:discussion}Discussion}

In this study, we used a neuromorphic chip to emulate a network, subject to plasticity, and showed a clear relation between criticality, task-performance and information theoretic fingerprint.
Most interestingly, simple tasks do not profit from criticality while complex ones do, showing that every task requires its own network state.

The state and hence computational properties can readily be tuned by changing the input strength, and thus a critical state can be reached without any parameter fine-tuning within the network.
This robust mechanism to adapt a network to task requirements is highly promising, especially for large networks where many parameters have to be tuned and in analog neuromorphic devices that are subject to noise in parameters and dynamics.

It has been generally suggested that criticality \textit{optimizes} task performance~\cite{munoz2018,boedecker2012}.
We show that this statement has to be specified: indeed, criticality \textit{maximizes} a number of properties, like the autocorrelation time (\cref{fig:corr}), the susceptibility (\cref{fig:variation}), as well as information theoretic measures (\cref{fig:memory_capacity,fig:info,fig:pidinternal}).
However, this maximization is apparently not at all necessary, potentially even detrimental, when dealing with simple tasks.
For our simple task, high network capacity results in maintenance of task-irrelevant information, and thereby harms performance.
This is underlined by our results that clearly show that all abstract computational properties are maximized at criticality, but only the complex tasks profit from criticality.
Hence, every task needs its own state and therefore a specific distance to critical dynamics.

The input strength could not only be controlled by changing $K_\mathrm{ext}$, the number of synapses of a neuron that were coupled to the input.
An equally valid choice is a change of external input \textit{rate} to each neuron.
In fact, we showed that changing the input rate has the same effects on the relation between criticality (figure~2 and~3 of Supplemental Material~\cite{supplement}), task performance, and information measures (figure~5 of Supplemental Material~\cite{supplement}) as changing $K_\mathrm{ext}$.
Moreover, in this framework the lowest input rates even allow to cross the critical point (figure~3 and~4 of Supplemental Material~\cite{supplement}).
Thus for both control mechanisms or a combination, there exists an optimal input strength, where the homeostatic mechanisms bring the network closest to critical.
This optimal input strength has been derived analytically for a mean-field network by Johannes Zierenberg~\cite{zierenberg2018}, and could potentially be used to predict the optimal input strength for other networks and tasks as well.

Not only the input strength, but also the strength of inhibition can act as a control parameter.
Inhibition plays a role in shaping collective dynamics and is known to generate oscillations~\cite{whittington2000,buzsaki2012}.
For a specific ratio of excitation and inhibition, criticality has been observed in neural networks~\cite{poil2012,shin2006,hesse2014,neto2017}.
Likewise, our networks has \SI{20}{\percent} inhibitory neurons.
However, inhibition would not be necessary for criticality~\cite{levina2007,wilting2018a}.  
Nevertheless, the existence of more than one control parameters (degree of input, input rate, and inhibition) allows for flexible adjustment even in cases where only one of them could be freely set without perturbing input coding.

Plasticity plays a central role in self-organization of network dynamics and computational properties.
In our model, the plasticity, neuron and synapse dynamics feature quite some level of biological detail (\cref{tab:parameters}), and thus results could potentially depend on them.
All synaptic weights are determined by the synaptic plasticity.
Here, we showed results for homeostasis and \gls{stdp} that implement the negative (anticausal) arm only.
When implementing the positive (causal) arm of \gls{stdp} in addition, the network destabilised, despite counteracting homeostasis.
This is a well known problem~\cite{keck2017}.
Our implementation is still similar to full \gls{stdp}, because anticaulsal correlations are weakened and the causal ones are indirectly strengthened by homeostasis.
With its similarity to \gls{stdp} and its inherent stability, our reduced implementation may be useful for future studies.

The characterization of the network in a task-dependent as well as in an task-independent manner is essential for understanding the impact of criticality on computation.
The computational properties in the vicinity of a critical point have been investigated by the classical measures AIS, I and TE alone~\cite{shew2013,mediano2017}, or by \gls{pid} alone ~\cite{tax2017,wibral2017}.
In this paper, we indeed showed that criticality maximizes capacity, but this does not necessarily translate to maximal task performance.
Moreover, the lagged I between the stimulus and the activity of neurons allows to estimate memory time-scales required to solve our tasks.
This enables us to understand how task complexity and the information-theoretic fingerprint are related. Such understanding is the basis for well-founded design decisions of future artificial architectures.

The presented framework is particularly useful for analog neuromorphic devices as analog components have inherent parameter noise as well as thermal noise, which potentially destabilize the network.
Here, the synaptic plasticity plays a key role in equalizing out particularly the parameter noise, as also demonstrated for \textit{short-term} plasticity~\cite{bill2010}, and thus makes knowledge about parameter variations, as well as specific calibration to some extend unnecessary.

Despite of the small system size ($N=32$ neurons only), the network not only showed signatures of criticality, but also developed quite complex computational capabilities, reflected in both, the task performance and the abstract information-theoretic quantities.
We expect that a scale-up of the system size would open even richer possibilities.
Such a scale-up would not even require fine-tuning of parameters, as the network self-tunes owing to the local-learning rules.
As soon as larger chips are available, we expect that the abilities of neuromorphic hardware could be exhausted in terms of speed and energy efficiency allowing for long, large-scale and powerful emulations.

Overall, we found a clear relation between criticality, task-performance and information theoretic fingerprint.
Our result contradicts the widespread statement that criticality is optimal for information processing in general:
While the distance to criticality clearly impacts performance on the reservoir task, we showed that only the complex tasks profit from criticality; for simple ones, criticality is detrimental. 
Mechanistically, the optimal working point for each task can be set very easily under homeostasis by adapting the mean input strength. 
This shows how critical phenomena can be harnessed in the design and optimization of artificial networks, and may explain why  biological neural networks operate not necessarily at criticality, but in the dynamically rich vicinity of a critical point, where they can tune their computational properties to task requirements~\cite{wilting2018c,wilting2019}. 

\section{\label{sec:methods}Methods}

We start with a description of the implemented network model, followed by a summary of the analysis techniques.
All parameters are listed in \cref{tab:parameters} and all variables in table~1 and~2 of the Supplemental Material~\cite{supplement}.

\begin{table}[t]
	\centering
	\begin{footnotesize}
    \begin{tabular}{lll}
        \toprule
    	\textbf{Parameter} 			        & \textbf{Symbol} 					& \textbf{Value} \\
    	\midrule
    	Threshold potential			        & $u_\mathrm{thresh}$				& \SI{554(21)}{\milli\volt} \\
    	Leak potential				        & $u_\mathrm{leak}$					& \SI{384(79)}{\milli\volt} \\
    	Reset potential				        & $u_\mathrm{reset}$				& \SI{319(18)}{\milli\volt} \\
    	Membrane capacitance 		        & $C_m$								& \SI{2.38(2)}{\nano\farad} \\
    	Membrane time constant		        & $\tau_\mathrm{mem}$				& \SI{1.6(10)}{\milli\second} \\
    	Refractory period			        & $\tau_\mathrm{ref}$				& \SI{4.9(5)}{\milli\second} \\
    	Synaptic time constant		        & $\tau_\mathrm{syn}^\mathrm{exc}$	& \SI{3.7(5)}{\milli\second} \\
    								        & $\tau_\mathrm{syn}^\mathrm{inh}$	& \SI{2.8(3)}{\milli\second} \\
    	Synaptic delay				        & $d_\mathrm{syn}$					& \SI{1.9(1)}{\milli\second} \\
    	Weight scaling   			        & $\gamma$          				& \SI{8.96(13)}{\micro\ampere} \\
    	Inhibitory synapses per neuron      & $N_\mathrm{inh}$                  & \SI{6}{} \\
    	Neurons                             & $N$                               & \SI{32}{} \\
    	Degree of input                     & $K_\mathrm{ext}$                  & \SI{6}{} - \SI{32}{} \\
    	Input rate                          & $\nu$                             & \SI{29}{\hertz} \\
    	\midrule
    	STDP time constant			        & $\tau_\mathrm{STDP}$				& \SI{6.8(12)}{\milli\second} \\
    	STDP amplitude 	                    & $\eta$							& \SI{0.071(23)}{} \\
    	Correlation scaling			        & $\lambda_\mathrm{stdp}$  			& \SI{11/128}{} \\
    	Drift parameter				        & $\lambda_\mathrm{drift}$ 			& \SI{1/512}{} \\
    	Range of random variable            & $n_\mathrm{amp}$					& \SI{15/16}{} \\
    	Bias of random variable		        & $\langle n \rangle$				& \SI{3/16}{} \\
    	\midrule
    	Burn-in experiment duration	        & $T^\mathrm{burnin}$				& \SI{625}{\second} \\
    	Static experiment duration          & $T^\mathrm{exp}$					& \SI{104}{\second} \\
    	Static trial experiment duration    & $T^\mathrm{static}$               & \SI{1}{\second} \\
    	Training experiment duration        & $T^\mathrm{train}$                & \SI{104}{\second} \\
    	Testing experiment duration         & $T^\mathrm{test}$                 & \SI{21}{\second} \\
    	Perturbation experiment duration    & $T^\mathrm{pert}$                 & \SI{2}{\second} \\
    	Perturbation time                   & $t_\mathrm{pert}$                 & \SI{1}{\second} \\
    	Initial weight				        & $w_{ij}^\mathrm{init}$			& \SI{0}{\micro\ampere} \\
    	Plasticity update period	        & $T$								& \SI{1}{\milli\second} \\ 
    	\midrule
    	Embedding dimension 		        & $l$								& 4 \\
    	Delays steps                        & $N_\tau$                          & 100 \\
    	\bottomrule
    \end{tabular}
    \end{footnotesize}
	\caption{%
		Overview of the model parameters.
		All time-constants are given in biological time.
		\Glsfirst{stdp} amplitudes as well as time constants where measured using 20 spike pairs.
		The errors indicate the standard deviation.}
	\label{tab:parameters}
\end{table}

\subsection{\label{subsec:Model}Model}

The results shown in this article are acquired on the mixed-signal neuromorphic hardware system described in~\cite{friedmann2017} (\cref{fig:chip}).
In the following a brief overview of the model, which is approximated by the physical implementation on the hardware, and the programmed plasticity rule is given.
Since the neuromorphic hardware system comprises analog electric circuits, transistor mismatch causes parameter fluctuations which can be compensated by calibration~\cite{pfeil2013,bruederle2011,neftci2010,neftci2011}.
Here, no explicit calibration on the basis of single neurons and synapses is applied.
Instead, only parameters common to all neurons/synapses are set such that all parts behave according to the listed equations, especially that all parts are sensible to input but silent in the absence of input.
This choice leads to uncertainties in the model parameters as reported in \cref{tab:parameters}.

\textbf{Neurons}: Implemented in analog circuitry, the neurons approximate current-based \gls{lif} neurons.
The membrane potential $u_j$ of the $j$-th neuron obeys:
\begin{equation}
\tau_\mathrm{mem} \frac{du_j}{dt} = - \left[u_j(t) - u_\mathrm{leak}\right] + \frac{I_j(t)}{g_\mathrm{leak}} \label{eq:membrane} \, ,
\end{equation}
with the membrane time constant $\tau_\mathrm{mem}$, the leak conductance $g_\mathrm{leak} = C_m / \tau_\mathrm{mem}$, the leak potential $u_\mathrm{leak}$ and the input current $I_j(t)$.
The $k$-th firing time of neuron $j$, $t_j^k$, is defined by a threshold criterion:
\begin{equation}
t_j^k : u_j(t_j^k) \geq u_\mathrm{thres}\, .
\end{equation}
Immediately after $t_j^k$, the membrane potential is clamped to the reset potential $u_j(t) = u_\mathrm{reset}$ for $t \in \left(t_j^k, t_j^k + \tau_\mathrm{ref}\right]$, with the refractory period  $\tau_\mathrm{ref}$.
The neuromorphic hardware system comprises $N = 32$ neurons, operating in continuous time due to the analog implementation.

\textbf{Synapses}: Like the membrane dynamics, the synapses are implemented in electrical circuits.
Each neuron features $N = 32$ presynaptic partners (in-degree is 32). 
The synaptic input currents onto the $j$-th neuron enter the neuronal dynamics in \cref{eq:membrane} as the sum of the input currents of all presynaptic partners $i$, $I_j(t) = \sum_{i=1}^N I_{ij}(t)$, where $I_{ij}(t)$ is given by:
\begin{align}
\tau_\mathrm{syn}^\mathrm{exc}\frac{\mathrm{d}I_{ij}(t)}{\mathrm{d}t} &= - I_{ij}(t) + I_{ij}^\text{ext}(t) + I_{ij}^\text{rec}(t)\, , \\
\tau_\mathrm{syn}^\mathrm{inh}\frac{\mathrm{d}I_{ij}(t)}{\mathrm{d}t} &= - I_{ij}(t) - I_{ij}^\text{ext}(t) - I_{ij}^\text{rec}(t)\, ,
\end{align}
with the excitatory and the inhibitory synaptic time constants $\tau_\mathrm{syn}^\mathrm{exc}$ and $\tau_\mathrm{syn}^\mathrm{inh}$.
$N_\mathrm{inh}$ synapses of every neuron $j$ are randomly selected to be inhibitory.
The external synaptic current $I_{ij}^\mathrm{ext}(t)$ depends on the $l$-th spike time of an external stimulus $i$, $s_i^l$, whereas the recurrent synaptic current $I_{ij}^\mathrm{rec}(t)$ depends on the $k$-th spike time of neuron $i$, $t_i^k$, each of which transmitted to neuron $j$:
\begin{align}
I_{ij}^\text{ext}(t) &= \sum_l \gamma\cdot w_{ij}^\mathrm{ext}\cdot\delta\left(t - s_i^l  - d_\mathrm{syn} \right) \, , \\
I_{ij}^\text{rec}(t) &= \sum_k \gamma\cdot w_{ij}^\mathrm{rec}\cdot\delta\left(t - t_i^k  - d_\mathrm{syn} \right)\, ,
\end{align}
with the synaptic delay $d_\mathrm{syn}$ and the weight conversion factor $\gamma$.
The synaptic weight from an external spike source $i$ to neuron $j$ is denoted by $w_{ij}^\mathrm{ext}$, and $w_{ij}^\mathrm{rec}$ is the synaptic weight from neuron $i$ to neuron $j$.
Every synapse either transmits external events $s_i^l$ or recurrent spikes $t_i^k$, i.e. if $w_{ij}^\mathrm{stim} \geq 0$ then $w_{ij}^\mathrm{rec} = 0$ and vice versa.

\textbf{Network}: The \gls{lif} neurons are potentially connected in an all-to-all fashion.
A randomly selected set of $K_\mathrm{ext}$ synapses of every neuron is chosen to be connected to the spike sources.
As every synapse could either transmit recurrent or input spikes, the $K_\mathrm{ext}$ synapses do not transmit recurrent spikes.

\textbf{Plasticity}: In the network, all synapses are plastic, the recurrent and the ones linked to the external input.
Therefore, we skip the superscript of the synaptic weight    and drop the distinction of $t_i^k$ and $s_i^k$ in the following description.
Weights are subject to three contributions: A weight drift controlled by the parameter $\lambda_\mathrm{drift}$, a correlation sensitive part controlled by $\lambda_\mathrm{stdp}$ and positively biased noise contributions.
This is very similar to \gls{stdp}, however with specific depression, but unspecific potentiation.
A specialized processor on the neuromorphic chip is programmed to update synaptic weights to $w_{ij}(t + T) = w_{ij}(t) + \Delta w_{ij}$ according to:
\begin{equation}
\Delta w_{ij} = \underbrace{-\lambda_\mathrm{stdp} f \left(t_i^k, t_j^l, t\right)}_{\text{specific depression}} - \underbrace{\lambda_\mathrm{drift} w_{ij}}_{\text{decay}} + \underbrace{n_{ij}(t)}_{\substack{\text{unspecific} \\ \text{potentiation}}} \, . \label{eq:update}
\end{equation}
The \gls{stdp}-kernel function $f$ depends on the pre- and postsynaptic spike times in the time interval $[t-T, t)$:
\begin{equation}
f\left(t_i^k, t_j^l, t\right) = \sum_{\substack{t_i^k, t_j^l}} \eta_\mathrm{stdp}\exp\left(\frac{t_j^l - t_i^k}{\tau_\mathrm{stdp}}\right) \label{eq:kernel} \, ,
\end{equation}
with $t_i^k > t_j^l$, and $t_i^k,t_j^l\in [t-T,t)$, and only nearest-neighbor spike times are considered in the sum~\cite{morrison2008}.
$\eta_\mathrm{stdp}$ and $\tau_\mathrm{stdp}$ denote the amplitude and the time constant of the \gls{stdp}-kernel.
The term $n_{ij}(t)$ adds a uniformly distributed, biased random variable:
\begin{equation}
n_{ij} \sim \mathrm{unif}\left( -n_\mathrm{amp}, n_\mathrm{amp} \right) + \langle n \rangle \, ,
\end{equation}
where $n_\mathrm{amp}$ specifies the range, while $\langle n \rangle$ is the bias of the random numbers.

The parameters $\lambda_\mathrm{stdp}$ and $\lambda_\mathrm{drift}$ are chosen such that the average combined force of the drift and the stochastic term is positive.
Thus, only the negative arm of \gls{stdp} is implemented.

\textbf{Initialization}: The synaptic weights are initialized to $w_{ij} = \SI{0}{\micro\ampere}$.
Afterwards, the network is stimulated by $N$ Poisson-distributed spiketrains of rate $\nu$ by the $K_\mathrm{ext}$ synapses of every neuron.
By applying \cref{eq:update} for the total duration $T^\mathrm{burnin}$ weights $w_{ij} \neq \SI{0}{\micro\ampere}$ develop.
For every $K_\mathrm{ext}$, the network is run 100 times, each with a different random seed.
If not stated otherwise, the resulting weight matrices are used as initial conditions for experiments with frozen weights ($\Delta w_{ij} = 0$) for a duration of $T^\mathrm{exp}$ on which the analysis is performed.

\textbf{Simulations}: To complement the hardware emulations, an idealized version of the network is implemented in \textit{Brian 2}~\cite{goodman2009}.
Specifically, no parameter or temporal noise is considered, and weights are not discretized as it is the case for the neuromorphic chip.
For simplicity, the degree of the input is implemented probabilistically by connecting each neuron-input pair with probability $K_\mathrm{ext} / N$ and each pair of neurons with probability $(N - K_\mathrm{ext}) / N$.

\subsection{\label{sec:evaluation}Evaluation}

\textbf{Binning}: The following measures rely on an estimate of activity, therefore we apply temporal binning:
\begin{equation}
    \tilde{x}_i(t) = \sum_k \mathds{1}\left(x_i^k \geq t\cdot \delta t, \, x_i^k < (t+1)\cdot \delta t\right) \, ,
\end{equation}
where $\delta t$ corresponds to the binwidth, and $\mathds{1}$ is the indicator function.
With this definition, we are able to define the binarized activity for a single process $i$:
\begin{equation}
x_i(t) = \min\left[1, \tilde{x}_i(t)\right] \label{eq:binned_single}\, .
\end{equation}
The variable $x_i(t)$ can represent either activity of a neuron in the network $a_i(t)$, or of a stimulus spike train $s_i(t)$, and correspondingly the spike times $x_i^k$ represent spikes of network neurons or stimulus (input) spike trains.

The population activity $a(t)$ of the network is defined as:
\begin{equation}
a(t) = \sum_{i=1}^N \tilde{a}_i(t) \, .
\end{equation}

\textbf{Neural avalanches}: A neural avalanche is a cascade of spikes in neural networks.
We extract avalanches from the population activity $a(t)$, obtained by binning the spike data with $\delta t$ corresponding to the mean inter-event interval, following established definitions.
In detail, one avalanche is separated from the subsequent one by at least one empty time bin~\cite{beggs2003}.
The size $s$ of an avalanche is defined as the number of spikes in consecutive non-empty time bins.
At criticality, the size distribution $P(s)$ is expected to follow a power-law.

To test for criticality, we compare whether a power-law or an exponential distributions fits the acquired avalanche distribution $P(s)$ better ~\cite{clauset2009}.
For the fitting, first the best matching distribution is determined based on the fit-likelihood.
The fit-range is fixed to $s\in\{4, 3\cdot N\}$ as the system is of finite-size.
An estimation of the critical exponent $\alpha_s$ and an exponential cutoff $s_\mathrm{cut}$ is obtained by fitting a truncated power-law:
\begin{equation}
\mathcal{P}_\mathrm{pl}(s) \propto s^{-\alpha_s} \exp{\left(-\frac{s}{s_\mathrm{cut}}\right)} \, ,
\label{eq:truncated_powerlaw}
\end{equation}
for $s\geq 1$.
Power law fits are performed with the Python package \textit{power-law} described in~\cite{alstott2014}.

\textbf{Fano-factor}: The variability of the population activity is quantified by the Fano factor $F = \sigma_a^2 / \mu_a$ where $\sigma_a^2$ is the variance and $\mu_a$ is the mean of the population activity $a(t)$, binned with $\delta t = \tau_\mathrm{ref}$.

\textbf{Trial-to-trial variability and susceptibility}: The trial-to-trial distance $\Delta_\mathrm{VRD}$ is obtained by stimulating  the same network twice with the same Poisson spike trains, leading to two different trials $m$ and $n$ influenced by variations caused by the physical implementation.
The resulting spike times in trial $m$ emitted by neuron $i$, termed $t_{i,m}^j$, are convolved with a Gaussian:
\begin{equation}
\tilde{t}_{i,m}(t) = \sum_j \int_0^{T^\mathrm{exp}} \exp{\left(-\frac{(t - t')^2}{2\sigma_\mathrm{VRD}^2}\right)} \delta(t' -{t}{_{i,m}^j}) \mathrm{dt'}\, ,
\end{equation}
and likewise for trial $n$.
The width is chosen to be $\sigma_\mathrm{VRD} = \tau_\mathrm{ref}$ and the temporal resolution for the integration is chosen to be \SI{0.1}{\milli\second}.
From different trials $m$ and $n$ the distance is calculated:
\begin{equation}
\Delta_\mathrm{VRD} = \frac{1}{\sigma_\mathrm{VRD}} \sum_{\substack{m,n\\m\neq n}}\sum_{i=1}^N \int_{-\infty}^\infty \frac{[\tilde{t}_{i,m}(t) - \tilde{t}_{i,n}(t)]^2}{[\tilde{t}_{i,m}(t) + \tilde{t}_{i,n}(t)]^2}\mathrm{dt} \, . \label{eq:distance}
\end{equation}

To obtain an estimate of the networks sensitivity $\chi$ to external perturbations, a pulse of $N_\mathrm{pert}$ additional spikes is embedded in the stimulating Poisson spike trains at time $t_\mathrm{pert}$:
\begin{equation}
\chi = \frac{a(t_\mathrm{pert} + \delta t) - a(t_\mathrm{pert})}{K_\mathrm{ext}^2}  \, , \label{eq:susceptibility}
\end{equation}
normalized to the number of external connection $K_\mathrm{ext}^2$ to compensate for the decoupling from external input with decreasing $K_\mathrm{ext}$.
The population activity is estimated with binsize $\delta t = d_\mathrm{syn}$.
By evaluating $\chi$ immediately after the perturbation, only the effect of the perturbation is captured by minimizing the impact of trial-to-trial variations.

To calculate $\chi$ and $\Delta_\mathrm{VRD}$, each weight matrix, obtained by the application of the plasticity rule, is used as initial condition for $10$ emulations with frozen weights and fixed seeds for the Poisson-distributed spike trains of duration $T^\mathrm{pert}$ and $T^\mathrm{static}$.
Additionally, a perturbation of size $N_\mathrm{pert}$ at $t_\mathrm{pert} = T^\mathrm{pert} / 2$ is embedded for the estimation of $\chi$.

\textbf{Autoregressive model}: Mathematically, the evolution of spiking neural networks is often approximated by a first-order autoregressive representation.
To assess the branching parameter $m$ of the network in analogy to~\cite{beggs2003,wilting2018a}, we make use of the following ansatz:
\begin{equation}
\langle a(t + 1) \rvert a(t) \rangle = m \cdot a(t) + h \, , \label{eq:ar}
\end{equation}
where the population activity in the next time step, $a(t+1)$ is determined by internal propagation within the network ($m$), and by external input $h$.
Here, $\langle . \rvert . \rangle $ denotes the conditional expectation and $m$ corresponds to the branching ratio.
For $m=1$, the system is critical, for $m>1$ the system is supercritical and activity grows exponentially on expectation (if not limited by finite size effects), whereas for $m<1$ the activity is stationary.
The branching parameter $m$ is linked to the autocorrelation time constant by $\tau_\mathrm{branch} = -\delta t / \ln{(m)}$.
To obtain the activity $a(t)$ the binwidth $\delta t$ is set to the refractory time $\tau_\mathrm{ref}$.
Estimating $m$ is straight forward here, as subsampling~\cite{priesemann2009,priesemann2014} does not impact the estimate.
Thus a classical estimator~\cite{wei1990} can be used, i.e. $m$ is equal to the linear regression between $a(t)$ and $a(t+1)$.
For model validation purposes, the autocorrelation function $\rho_{a,a}$ is calculated on the population activity $a(t)$ binned with $\delta t = \tau_\mathrm{ref}$:
\begin{equation}
\rho_{a,a}(t') = \frac{1}{\sigma_a^2} \sum_{t=1}^{T^\mathrm{exp} / \delta t - t'} (a(t) - \mu_a) (a(t + t') - \mu_a) \, ,
\end{equation}
where $\sigma_a$ is the standard deviation, and $\mu_a$ the mean of the population activity.
Subsequently, $\rho_{a,a}$ is fitted by an exponential to yield the time constant $\tau_\mathrm{corr}$.

\textbf{Information theory}: We use notation, concepts and definitions as outlined in the review~\cite{wibral2015}. In brief, the time series produced by two neurons represent two stationary random processes $X_1$ and $X_2$, composed of random variables $X_1(t)$ and $X_2(t)$, $t=1,...,n$, with realizations $x_1(t)$ and $x_2(t)$.
The corresponding embedding vectors are given in bold font, e.g $\mathbf{X_1^l}(t) = \{X_1(t), X_1(t-1), ..., X_1(t - l + 1)\}$.
The embedding vector $\mathbf{X_1^l}(t)$ is constructed such that it renders the variable $X_1(t+1)$ conditionally independent of all random variables $X_1(t')$ with $t'<t-l+1$, i.e. $p(X_1(t+1) \rvert \mathbf{X_1^l}(t), X_1(t')) = p(X_1(t+1) \rvert \mathbf{X_1^l}(t))$.
Here, $(\cdot \vert \cdot)$ denotes the conditional.

The \glsfirst{h} and \glsfirst{i} are calculated for the random variables $X_1$, $X_2$, if not denoted otherwise.
This is equivalent to using $l=1$ above, e.g. $\mathrm{H}(X_1)$ and $\mathrm{I}(X_1: X_2) = \mathrm{H}(X_1) - \mathrm{H}(X_1 \rvert X_2)$.
We abbreviate the past state of spike train $1$ by $\mathbf{X_1^-}$: thus $\mathbf{X_1^l}(t-1) = \{X_1(t-1), X(t-2),..., X_1(t-l)\}$.
The current value of the spike train is denoted by $X_1$.
With this notation the \glsfirst{ais} of e.g. $X_1$ is given by:
\begin{equation}
\mathrm{AIS}(X_1) = \mathrm{I}(X_1: \mathbf{X_1^-})\, .
\end{equation}
In the same way, we define the \glsfirst{te} between source $X_1$ and target $X_2$:
\begin{equation}
\mathrm{TE}(X_1 \rightarrow X_2) = \mathrm{I}(X_2: \mathbf{X_1^-} \rvert \mathbf{X_2^-})\, .
\end{equation}

The lagged mutual information for time lag $\tau$ is defined as $\mathrm{I}_\tau(X_1: X_2) = \mathrm{I}_\tau\left(X_1(t): X_2(t+\tau)\right)$ is estimated. 
Integrating the lagged $\mathrm{I}$ defines the \glsfirst{mc}:
\begin{equation}
\mathrm{MC}(X_1: X_2) = \sum_{\tau = 1}^{N_\tau} \delta t\left[\mathrm{I}_\tau(X_1: X_2) - \mathrm{I}_{N_\tau}(X_1: X_2)\right] \label{eq:mc} \, ,
\end{equation}
with a maximal delay $N_\tau = 100$.
The \gls{i} of a sufficiently large at $N_\tau$ is subtracted to account for potential estimation biases.

To access the \textit{information modification} the novel concept of \gls{pid} is applied~\cite{williams2010,bertschinger2013,wibral2017}.
Intuitively, information modification in a pairwise consideration should correspond to the information about the present state of a process only available when considering both, the own process past and the past of a source process.
Therefore, the joint mutual information $\mathrm{I}(X_1: \mathbf{X_1^-}, \mathbf{X_2^-})$ is decomposed by \gls{pid} into the unique, shared (redundant), and synergistic contributions to the future spiking of one neuron, $X_1$, from its own past $\mathbf{X_1^-}$, and the past of a second neuron or an input stimulus $\mathbf{X_2^-}$:
In more detail, we quantify:
\begin{enumerate}
\item The unique information $\mathrm{I}_\mathrm{unq}(X_1: \mathbf{X_1^-} \setminus \mathbf{X_2^-})$ which is contributed from the neurons own past.
\item The unique information $\mathrm{I}_\mathrm{unq}(X_1: \mathbf{X_2^-} \setminus \mathbf{X_1^-})$ that is contributed from a different spike train (neuron or stimulus).
\item The shared information $\mathrm{I}_\mathrm{shd}(X_1: \mathbf{X_2^-}; \mathbf{X_1^-})$ which describes the redundant contribution.
\item The synergistic information $\mathrm{I}_\mathrm{syn}(X_1: \mathbf{X_2^-}; \mathbf{X_1^-})$, i.e. the information that can only be obtained when having knowledge about \emph{both} past states.
\end{enumerate}
$\mathrm{I}_\mathrm{syn}$ is what we consider to be a suitable measure for information modification~\cite{wibral2017}.

The joint mutual information as defined here is the sum of the \gls{ais} and the \gls{te}:
\begin{equation}
\mathrm{I}(X_1: \mathbf{X_1^-}, \mathbf{X_2^-}) = \mathrm{I}(X_1: \mathbf{X_1^-}) + \mathrm{I}(X_1 : \mathbf{X_2^-} \rvert \mathbf{X_1^-}) \, .
\end{equation}

We calculated \Gls{h}, \gls{ais}, \gls{i} and \gls{te} with the toolbox \textsl{JIDT}~\cite{lizier2014}, whereas the \gls{pid} was estimated with the \textsl{BROJA-2PID} estimator~\cite{makkeh2018}.
The activity is obtained by binning the spike data with $\delta t = \tau_\mathrm{ref}$ and setting $l$ to $4$ to incorporate sufficient history.
\Gls{i} and \gls{te} as well as the \gls{pid} were calculated pairwise between all possible combinations of processes.
Results are typically normalized by \gls{h} to compensate for potential changes in the firing rate for changing values of $K_\mathrm{ext}$ (figure~1 of Supplemental Material~\cite{supplement}).
For the pairwise measures, \gls{h} of the target neuron is used for normalization.

\textbf{Resevoir computing}: The performance of the neural network as a reservoir~\cite{maass2002,jaeger2001} is quantified using a variant of the $n$-bit parity task.
The network weights are frozen (i.e. plasticity is disabled) to ensure that the network state is not changed by the input.

To solve the parity requires to classify from the network activity $a_j(t)$, whether the last $n$ bits of input carried an odd or even number of spikes.
The network is stimulated with a single Poisson-distributed spike train of frequency $\nu$ acting equally on all external synapses, i.e. the input spike times are $s_i^k = s^k\,\forall\,i$. 
Spike times are binned according to \cref{eq:binned_single} with binwidth $\delta t$ to get a measure of the $n$ past bits.
The resulting stimulus activity $s(t)$ is used to calculate the $n$-bit \textbf{parity} function according to:
\begin{equation}
    p_n\left[s(t)\right]= s(t) \oplus s(t-1) \oplus ... \oplus s(t-n+1) \, ,
\end{equation}
with $p_n\left[s(t)\right] \in \{0,1\}$ and the modulus $2$ addition $\oplus$, i.e. whether an odd or even number of spikes occurred in the $n$ past time steps of duration $\delta t$.

On the activity $a_j(t)$ of a randomly selected subset $\mathcal{U}$ of neurons with cardinality $N_\mathrm{read}$ a classifier is trained:
\begin{equation}
	v(t) = \Theta\left(\sum_{j\in\mathcal{U}} w_{j}a_j(t) - \frac{1}{2}\right) \, ,
\end{equation}
where $\Theta(\cdot)$ is the Heaviside function, and $v(t)$ is the predicted label.
The weight vector $w_{j}$ of the classifier is determined using linear regression on a set of training data $s_\mathrm{train}$ of duration $T^\mathrm{train}$:
\begin{equation}
    w_j = \argmin_{w_j} \left(\sum_{t=0}^{T^\mathrm{train}/\delta t-1} \left\vert p_n\left[s_\mathrm{train}(t)\right] - w_j a_j(t) \right\vert^2\right) \label{eq:regression} \, .
\end{equation}
The network's performance on the parity task is quantified by $\mathrm{I}\left(p_n\left[s_\mathrm{test}(t)\right], v(t)\right)$ on a test data set $s_\mathrm{test}$ of duration $T^\mathrm{test}$.
The performance $\mathrm{I}$ is offset corrected by training the very same classifier on a shuffled version of $p_n[s(t)]$.
Moreover, we weighted each sample in the regression in \cref{eq:regression} with the relative occurrence of their respective class to compensate for imbalance.
Temporal binning with $\delta t = \SI{1}{\milli\second}$ is applied to $s_\mathrm{train}$, $s_\mathrm{test}$ as well as $a_j(t)$.

In a second task, the stimulus activity $s(t)$ is used to calculate the $n$-bit sum according to:
\begin{equation}
    z_n\left[s(t)\right] = s(t) + s(t-1) + ... + s(t-n+1) \, ,
\end{equation}
i.e. how many spikes occurred in the $n$ past time steps of duration $\delta t$.
Here, the classifier described above is extended to multiple classes by adding readout units.
The decision of the classifier is implemented by a winner-take-all mechanism across units.

\section*{Acknowledgments}

This work has received funding from the European Union Sixth Framework Programme ([FP6/2002-2006]) under grant agreement no 15879 (FACETS), the European Union Seventh Framework Programme ([FP7/2007-2013]) under grant agreement no 604102 (HBP), 269921 (BrainScaleS) and 243914 (Brain-i-Nets) and the Horizon 2020 Framework Programme ([H2020/2014-2020]) under grant agreement no 720270  and 785907 (HBP), as well as the Manfred St\"{a}rk Foundation.
Viola Priesemann was supported by the Max Planck Society.
Michael Wibral has received funding from the VolkswagenStiftung.
The authors acknowledge support by the state of Baden-W\"{u}rttemberg through bwHPC.
We thank Matthias Loidold, Joao Pinheiro Neto, Korbinian Schreiber, Paul Spitzner, and Johannes Zierenberg for helpful comments on the manuscript.

\section*{Author Contributions}

B.C, D.S., and V.P. conceptualized the work.
B.C. and V.P. wrote the manuscript.
B.C. conducted the experiments; D.S. contributed to many of the experiments; M.K. contributed to the parity task.
M.W. contributed to the information theory.
V.P. supervised the work.
J.S. is the architect and lead designer of the neuromorphic platform.
J.S. and K.M. provided conceptual and scientific advice.


\begin{thebibliography}{81}
\providecommand{\natexlab}[1]{#1}
\providecommand{\url}[1]{\texttt{#1}}
\expandafter\ifx\csname urlstyle\endcsname\relax
  \providecommand{\doi}[1]{doi: #1}\else
  \providecommand{\doi}{doi: \begingroup \urlstyle{rm}\Url}\fi

\bibitem[Boedecker et~al.(2012)Boedecker, Obst, Lizier, Mayer, and
  Asada]{boedecker2012}
Joschka Boedecker, Oliver Obst, Joseph~T Lizier, N~Michael Mayer, and Minoru
  Asada.
\newblock Information processing in echo state networks at the edge of chaos.
\newblock \emph{Theory in Biosciences}, 131\penalty0 (3):\penalty0 205--213,
  2012.

\bibitem[Bertschinger and Natschl{\"a}ger(2004)]{bertschinger2004}
Nils Bertschinger and Thomas Natschl{\"a}ger.
\newblock Real-time computation at the edge of chaos in recurrent neural
  networks.
\newblock \emph{Neural computation}, 16\penalty0 (7):\penalty0 1413--1436,
  2004.

\bibitem[Legenstein and Maass(2007)]{legenstein2007}
Robert Legenstein and Wolfgang Maass.
\newblock Edge of chaos and prediction of computational performance for neural
  circuit models.
\newblock \emph{Neural Networks}, 20\penalty0 (3):\penalty0 323--334, 2007.

\bibitem[Kinouchi and Copelli(2006)]{kinouchi2006}
Osame Kinouchi and Mauro Copelli.
\newblock Optimal dynamical range of excitable networks at criticality.
\newblock \emph{Nature physics}, 2\penalty0 (5):\penalty0 348--351, 2006.

\bibitem[Shew and Plenz(2013)]{shew2013}
Woodrow~L Shew and Dietmar Plenz.
\newblock The functional benefits of criticality in the cortex.
\newblock \emph{The neuroscientist}, 19\penalty0 (1):\penalty0 88--100, 2013.

\bibitem[Del~Papa et~al.(2017)Del~Papa, Priesemann, and Triesch]{delpapa2017}
Bruno Del~Papa, Viola Priesemann, and Jochen Triesch.
\newblock Criticality meets learning: Criticality signatures in a
  self-organizing recurrent neural network.
\newblock \emph{PloS one}, 12\penalty0 (5):\penalty0 e0178683, 2017.

\bibitem[Langton(1990)]{langton1990}
Chris~G Langton.
\newblock Computation at the edge of chaos: phase transitions and emergent
  computation.
\newblock \emph{Physica D: Nonlinear Phenomena}, 42\penalty0 (1-3):\penalty0
  12--37, 1990.

\bibitem[Yam and Chow(2000)]{yam2000}
Jim~YF Yam and Tommy~WS Chow.
\newblock A weight initialization method for improving training speed in
  feedforward neural network.
\newblock \emph{Neurocomputing}, 30\penalty0 (1-4):\penalty0 219--232, 2000.

\bibitem[Thimm and Fiesler(1995)]{thimm1995}
Georg Thimm and Emile Fiesler.
\newblock Neural network initialization.
\newblock In \emph{International Workshop on Artificial Neural Networks}, pages
  535--542. Springer, 1995.

\bibitem[Goodfellow et~al.(2016)Goodfellow, Bengio, and
  Courville]{goodfellow2016}
Ian Goodfellow, Yoshua Bengio, and Aaron Courville.
\newblock \emph{Deep Learning}.
\newblock MIT Press, 2016.

\bibitem[Harris(2002)]{harris2002}
Theodore~E Harris.
\newblock \emph{The theory of branching processes}.
\newblock Courier Corporation, 2002.

\bibitem[Munoz(2018{\natexlab{a}})]{munoz2017}
Miguel~A Munoz.
\newblock Colloquium: Criticality and dynamical scaling in living systems.
\newblock \emph{Reviews of Modern Physics}, 90\penalty0 (3):\penalty0 031001,
  2018{\natexlab{a}}.

\bibitem[Wilting et~al.(2018)Wilting, Dehning, Neto, Rudelt, Wibral,
  Zierenberg, and Priesemann]{wilting2018c}
Jens Wilting, Jonas Dehning, Joao~Pinheiro Neto, Lucas Rudelt, Michael Wibral,
  Johannes Zierenberg, and Viola Priesemann.
\newblock Operating in a reverberating regime enables rapid tuning of network
  states to task requirements.
\newblock \emph{Frontiers in systems neuroscience}, 12, 2018.

\bibitem[Barnett et~al.(2013)Barnett, Lizier, Harr{\'e}, Seth, and
  Bossomaier]{barnett2013}
Lionel Barnett, Joseph~T Lizier, Michael Harr{\'e}, Anil~K Seth, and Terry
  Bossomaier.
\newblock Information flow in a kinetic ising model peaks in the disordered
  phase.
\newblock \emph{Physical review letters}, 111\penalty0 (17):\penalty0 177203,
  2013.

\bibitem[Tka{\v{c}}ik et~al.(2015)Tka{\v{c}}ik, Mora, Marre, Amodei, Palmer,
  Berry, and Bialek]{tkavcik2015}
Ga{\v{s}}per Tka{\v{c}}ik, Thierry Mora, Olivier Marre, Dario Amodei,
  Stephanie~E Palmer, Michael~J Berry, and William Bialek.
\newblock Thermodynamics and signatures of criticality in a network of neurons.
\newblock \emph{Proceedings of the National Academy of Sciences}, 112\penalty0
  (37):\penalty0 11508--11513, 2015.

\bibitem[Munoz(2018{\natexlab{b}})]{munoz2018}
Miguel~A Munoz.
\newblock Colloquium: Criticality and dynamical scaling in living systems.
\newblock \emph{Reviews of Modern Physics}, 90\penalty0 (3):\penalty0 031001,
  2018{\natexlab{b}}.

\bibitem[Levina et~al.(2007)Levina, Herrmann, and Geisel]{levina2007}
Anna Levina, J~Michael Herrmann, and Theo Geisel.
\newblock Dynamical synapses causing self-organized criticality in neural
  networks.
\newblock \emph{Nature physics}, 3\penalty0 (12):\penalty0 857--860, 2007.

\bibitem[Meisel and Gross(2009)]{meisel2009}
Christian Meisel and Thilo Gross.
\newblock Adaptive self-organization in a realistic neural network model.
\newblock \emph{Physical Review E}, 80\penalty0 (6):\penalty0 061917, 2009.

\bibitem[Stepp et~al.(2015)Stepp, Plenz, and Srinivasa]{stepp2015}
Nigel Stepp, Dietmar Plenz, and Narayan Srinivasa.
\newblock Synaptic plasticity enables adaptive self-tuning critical networks.
\newblock \emph{PLoS computational biology}, 11\penalty0 (1):\penalty0
  e1004043, 2015.

\bibitem[de~Andrade~Costa et~al.(2015)de~Andrade~Costa, Copelli, and
  Kinouchi]{andrade2015}
Ariadne de~Andrade~Costa, Mauro Copelli, and Osame Kinouchi.
\newblock Can dynamical synapses produce true self-organized criticality?
\newblock \emph{Journal of Statistical Mechanics: Theory and Experiment},
  2015\penalty0 (6):\penalty0 P06004, 2015.

\bibitem[Tetzlaff et~al.(2010)Tetzlaff, Okujeni, Egert, W{\"o}rg{\"o}tter, and
  Butz]{tetzlaff2010}
Christian Tetzlaff, Samora Okujeni, Ulrich Egert, Florentin W{\"o}rg{\"o}tter,
  and Markus Butz.
\newblock Self-organized criticality in developing neuronal networks.
\newblock \emph{PLoS computational biology}, 6\penalty0 (12):\penalty0
  e1001013, 2010.

\bibitem[Zierenberg et~al.(2018)Zierenberg, Wilting, and
  Priesemann]{zierenberg2018}
Johannes Zierenberg, Jens Wilting, and Viola Priesemann.
\newblock Homeostatic plasticity and external input shape neural network
  dynamics.
\newblock \emph{Phys. Rev. X}, 8:\penalty0 031018, Jul 2018.
\newblock \doi{10.1103/PhysRevX.8.031018}.

\bibitem[Poil et~al.(2012)Poil, Hardstone, Mansvelder, and
  Linkenkaer-Hansen]{poil2012}
Simon-Shlomo Poil, Richard Hardstone, Huibert~D Mansvelder, and Klaus
  Linkenkaer-Hansen.
\newblock Critical-state dynamics of avalanches and oscillations jointly emerge
  from balanced excitation/inhibition in neuronal networks.
\newblock \emph{Journal of Neuroscience}, 32\penalty0 (29):\penalty0
  9817--9823, 2012.

\bibitem[Shin and Kim(2006)]{shin2006}
Chang-Woo Shin and Seunghwan Kim.
\newblock Self-organized criticality and scale-free properties in emergent
  functional neural networks.
\newblock \emph{Physical Review E}, 74\penalty0 (4):\penalty0 045101, 2006.

\bibitem[Hebb(1949)]{hebb1949}
Donald~Olding Hebb.
\newblock \emph{The organization of behavior: A neuropsychological theory}.
\newblock John Wiley \& Sons, 1949.

\bibitem[Hopfield(1982)]{hopfield1982}
John~J Hopfield.
\newblock Neural networks and physical systems with emergent collective
  computational abilities.
\newblock \emph{Proceedings of the national academy of sciences}, 79\penalty0
  (8):\penalty0 2554--2558, 1982.

\bibitem[Bi and Poo(1998)]{bi1998}
{Guo-qiang} Bi and {Mu-ming} Poo.
\newblock Synaptic modifications in cultured hippocampal neurons: dependence on
  spike timing, synaptic strength, and postsynaptic cell type.
\newblock \emph{Journal of neuroscience}, 18\penalty0 (24):\penalty0
  10464--10472, October 1998.

\bibitem[Markram et~al.(1997)Markram, L{\"u}bke, Frotscher, and
  Sakmann]{markram1997}
Henry Markram, Joachim L{\"u}bke, Michael Frotscher, and Bert Sakmann.
\newblock Regulation of synaptic efficacy by coincidence of postsynaptic {AP}s
  and {EPSP}s.
\newblock \emph{Science}, 275\penalty0 (5297):\penalty0 213--215, Janurary
  1997.

\bibitem[Loidolt et~al.(forthcoming)Loidolt, Rudelt, and
  Priesemann]{loidolt_in_prep}
Matthias Loidolt, Lucas Rudelt, and Viola Priesemann.
\newblock Noise input enhances memory performance in learning recurrent
  networks.
\newblock \emph{in preparation}, forthcoming.

\bibitem[Wibral et~al.(2015)Wibral, Lizier, and Priesemann]{wibral2015}
Michael Wibral, Joseph~T Lizier, and Viola Priesemann.
\newblock Bits from brains for biologically inspired computing.
\newblock \emph{Frontiers in Robotics and AI}, 2:\penalty0 5, 2015.

\bibitem[Shannon(1948)]{shannon1948}
C.~E. Shannon.
\newblock A mathematical theory of communication.
\newblock \emph{The Bell System Technical Journal}, 27\penalty0 (3):\penalty0
  379--423, July 1948.
\newblock ISSN 0005-8580.
\newblock \doi{10.1002/j.1538-7305.1948.tb01338.x}.

\bibitem[Cover and Thomas(2012)]{cover2012}
Thomas~M Cover and Joy~A Thomas.
\newblock \emph{Elements of information theory}.
\newblock John Wiley \& Sons, 2012.

\bibitem[Williams and Beer(2010)]{williams2010}
PL~Williams and RD~Beer.
\newblock Decomposing multivariate information.
\newblock \emph{arXiv preprint arXiv:1004.2515}, 2010.

\bibitem[Bertschinger et~al.(2013)Bertschinger, Rauh, Olbrich, and
  Jost]{bertschinger2013}
Nils Bertschinger, Johannes Rauh, Eckehard Olbrich, and J{\"u}rgen Jost.
\newblock Shared information—new insights and problems in decomposing
  information in complex systems.
\newblock In \emph{Proceedings of the European conference on complex systems
  2012}, pages 251--269. Springer, 2013.

\bibitem[Lizier et~al.(2018)Lizier, Bertschinger, Jost, and Wibral]{lizier2018}
Joseph~T Lizier, Nils Bertschinger, J{\"u}rgen Jost, and Michael Wibral.
\newblock Information decomposition of target effects from multi-source
  interactions: Perspectives on previous, current and future work.
\newblock \emph{Entropy}, 20\penalty0 (4):\penalty0 307, 2018.

\bibitem[Wibral et~al.(2017{\natexlab{a}})Wibral, Finn, Wollstadt, Lizier, and
  Priesemann]{wibral2017}
Michael Wibral, Conor Finn, Patricia Wollstadt, Joseph~T. Lizier, and Viola
  Priesemann.
\newblock Quantifying information modification in developing neural networks
  via partial information decomposition.
\newblock \emph{Entropy}, 19\penalty0 (9), 2017{\natexlab{a}}.
\newblock ISSN 1099-4300.
\newblock \doi{10.3390/e19090494}.

\bibitem[Wibral et~al.(2017{\natexlab{b}})Wibral, Priesemann, Kay, Lizier, and
  Phillips]{wibral2017b}
Michael Wibral, Viola Priesemann, Jim~W Kay, Joseph~T Lizier, and William~A
  Phillips.
\newblock Partial information decomposition as a unified approach to the
  specification of neural goal functions.
\newblock \emph{Brain and cognition}, 112:\penalty0 25--38, 2017{\natexlab{b}}.

\bibitem[Mead(1990)]{mead1990}
Carver Mead.
\newblock Neuromorphic electronic systems.
\newblock \emph{Proceedings of the IEEE}, 78\penalty0 (10):\penalty0
  1629--1636, October 1990.

\bibitem[Douglas et~al.(1995)Douglas, Mahowald, and Mead]{douglas1995}
Rodney Douglas, Misha Mahowald, and Carver Mead.
\newblock Neuromorphic analogue {VLSI}.
\newblock \emph{Annual review of neuroscience}, 18\penalty0 (1):\penalty0
  255--281, 1995.

\bibitem[Aamir et~al.(2018)Aamir, Stradmann, M{\"u}ller, Pehle, Hartel,
  Gr{\"u}bl, Schemmel, and Meier]{aamir2018}
Syed~Ahmed Aamir, Yannik Stradmann, Paul M{\"u}ller, Christian Pehle, Andreas
  Hartel, Andreas Gr{\"u}bl, Johannes Schemmel, and Karlheinz Meier.
\newblock An accelerated lif neuronal network array for a large-scale
  mixed-signal neuromorphic architecture.
\newblock \emph{IEEE Transactions on Circuits and Systems I: Regular Papers},
  65\penalty0 (12):\penalty0 4299--4312, 2018.

\bibitem[Friedmann et~al.(2017)Friedmann, Schemmel, Gr{\"u}bl, Hartel, Hock,
  and Meier]{friedmann2017}
S.~Friedmann, J.~Schemmel, A.~Gr{\"u}bl, A.~Hartel, M.~Hock, and K.~Meier.
\newblock Demonstrating hybrid learning in a flexible neuromorphic hardware
  system.
\newblock \emph{IEEE Transactions on Biomedical Circuits and Systems},
  11\penalty0 (1):\penalty0 128--142, Feb 2017.
\newblock ISSN 1932-4545.
\newblock \doi{10.1109/TBCAS.2016.2579164}.

\bibitem[sup()]{supplement}
See supplemental material for controlling the input strength by variation of
  the input frequency.

\bibitem[Wunderlich et~al.(2019)Wunderlich, Kungl, M{\"u}ller, Hartel,
  Stradmann, Aamir, Gr{\"u}bl, Heimbrecht, Schreiber, St{\"o}ckel,
  et~al.]{wunderlich2019}
Timo Wunderlich, Akos~Ferenc Kungl, Eric M{\"u}ller, Andreas Hartel, Yannik
  Stradmann, Syed~Ahmed Aamir, Andreas Gr{\"u}bl, Arthur Heimbrecht, Korbinian
  Schreiber, David St{\"o}ckel, et~al.
\newblock Demonstrating advantages of neuromorphic computation: a pilot study.
\newblock \emph{Frontiers in Neuroscience}, 13:\penalty0 260, 2019.

\bibitem[Zapperi et~al.(1995)Zapperi, Lauritsen, and Stanley]{zapperi1995}
Stefano Zapperi, Kent~B{\ae}kgaard Lauritsen, and H~Eugene Stanley.
\newblock Self-organized branching processes: mean-field theory for avalanches.
\newblock \emph{Physical review letters}, 75\penalty0 (22):\penalty0 4071,
  1995.

\bibitem[Watson and Galton(1875)]{watson1875}
Henry~William Watson and Francis Galton.
\newblock On the probability of the extinction of families.
\newblock \emph{The Journal of the Anthropological Institute of Great Britain
  and Ireland}, 4:\penalty0 138--144, 1875.

\bibitem[Wilting and Priesemann(2018)]{wilting2018a}
Jens Wilting and Viola Priesemann.
\newblock Inferring collective dynamical states from widely unobserved systems.
\newblock \emph{Nature Communications}, 9\penalty0 (1):\penalty0 2325, 2018.
\newblock ISSN 2041-1723.
\newblock \doi{10.1038/s41467-018-04725-4}.

\bibitem[Beggs and Plenz(2003)]{beggs2003}
John~M Beggs and Dietmar Plenz.
\newblock Neuronal avalanches in neocortical circuits.
\newblock \emph{Journal of neuroscience}, 23\penalty0 (35):\penalty0
  11167--11177, 2003.

\bibitem[Priesemann and Shriki(2018)]{priesemann2018can}
Viola Priesemann and Oren Shriki.
\newblock Can a time varying external drive give rise to apparent criticality
  in neural systems?
\newblock \emph{PLoS computational biology}, 14\penalty0 (5):\penalty0
  e1006081, 2018.

\bibitem[Clauset et~al.(2009)Clauset, Shalizi, and Newman]{clauset2009}
Aaron Clauset, Cosma~Rohilla Shalizi, and Mark~EJ Newman.
\newblock Power-law distributions in empirical data.
\newblock \emph{SIAM review}, 51\penalty0 (4):\penalty0 661--703, 2009.

\bibitem[Shew et~al.(2011)Shew, Yang, Yu, Roy, and Plenz]{shew2011}
Woodrow~L Shew, Hongdian Yang, Shan Yu, Rajarshi Roy, and Dietmar Plenz.
\newblock Information capacity and transmission are maximized in balanced
  cortical networks with neuronal avalanches.
\newblock \emph{Journal of neuroscience}, 31\penalty0 (1):\penalty0 55--63,
  2011.

\bibitem[Maass et~al.(2002)Maass, Natschl{\"a}ger, and Markram]{maass2002}
Wolfgang Maass, Thomas Natschl{\"a}ger, and Henry Markram.
\newblock Real-time computing without stable states: A new framework for neural
  computation based on perturbations.
\newblock \emph{Neural computation}, 14\penalty0 (11):\penalty0 2531--2560,
  2002.

\bibitem[Jaeger(2001)]{jaeger2001}
Herbert Jaeger.
\newblock The “echo state” approach to analysing and training recurrent
  neural networks-with an erratum note.
\newblock \emph{Bonn, Germany: German National Research Center for Information
  Technology GMD Technical Report}, 148\penalty0 (34):\penalty0 13, 2001.

\bibitem[Sch{\"u}rmann et~al.(2005)Sch{\"u}rmann, Meier, and
  Schemmel]{schurmann2005}
Felix Sch{\"u}rmann, Karlheinz Meier, and Johannes Schemmel.
\newblock Edge of chaos computation in mixed-mode vlsi-a hard liquid.
\newblock In \emph{Advances in neural information processing systems}, pages
  1201--1208, 2005.

\bibitem[Schemmel et~al.(2010)Schemmel, Briiderle, Griibl, Hock, Meier, and
  Millner]{schemmel2010}
Johannes Schemmel, Daniel Briiderle, Andreas Griibl, Matthias Hock, Karlheinz
  Meier, and Sebastian Millner.
\newblock A wafer-scale neuromorphic hardware system for large-scale neural
  modeling.
\newblock In \emph{Proceedings of 2010 IEEE International Symposium on Circuits
  and Systems}, pages 1947--1950. IEEE, 2010.

\bibitem[Brette(2019)]{brette2019}
Romain Brette.
\newblock Is coding a relevant metaphor for the brain?
\newblock \emph{Behavioral and Brain Sciences}, 42, 2019.

\bibitem[Bernardi and Lindner(2017)]{bernardi2017}
Davide Bernardi and Benjamin Lindner.
\newblock Optimal detection of a localized perturbation in random networks of
  integrate-and-fire neurons.
\newblock \emph{Physical review letters}, 118\penalty0 (26):\penalty0 268301,
  2017.

\bibitem[Zierenberg et~al.(2019)Zierenberg, Wilting, Priesemann, and
  Levina]{zierenberg2019}
Johannes Zierenberg, Jens Wilting, Viola Priesemann, and Anna Levina.
\newblock Tailored ensembles of neural networks optimize sensitivity to
  stimulus statistics.
\newblock \emph{arXiv preprint arXiv:1905.10401}, 2019.

\bibitem[Lizier et~al.(2008)Lizier, Prokopenko, and Zomaya]{Lizier2008}
Joseph~T Lizier, Mikhail Prokopenko, and Albert~Y Zomaya.
\newblock The information dynamics of phase transitions in random boolean
  networks.
\newblock In \emph{ALIFE}, pages 374--381, 2008.

\bibitem[Wibral et~al.(2014)Wibral, Lizier, V{\"o}gler, Priesemann, and
  Galuske]{wibral2014}
Michael Wibral, Joseph Lizier, Sebastian V{\"o}gler, Viola Priesemann, and Ralf
  Galuske.
\newblock Local active information storage as a tool to understand distributed
  neural information processing.
\newblock \emph{Frontiers in neuroinformatics}, 8:\penalty0 1, 2014.

\bibitem[Schreiber(2000)]{schreiber2000}
Thomas Schreiber.
\newblock Measuring information transfer.
\newblock \emph{Physical review letters}, 85\penalty0 (2):\penalty0 461, 2000.

\bibitem[Whittington et~al.(2000)Whittington, Traub, Kopell, Ermentrout, and
  Buhl]{whittington2000}
Miles~A Whittington, RD~Traub, N~Kopell, B~Ermentrout, and EH~Buhl.
\newblock Inhibition-based rhythms: experimental and mathematical observations
  on network dynamics.
\newblock \emph{International journal of psychophysiology}, 38\penalty0
  (3):\penalty0 315--336, 2000.

\bibitem[Buzs{\'a}ki and Wang(2012)]{buzsaki2012}
Gy{\"o}rgy Buzs{\'a}ki and Xiao-Jing Wang.
\newblock Mechanisms of gamma oscillations.
\newblock \emph{Annual review of neuroscience}, 35:\penalty0 203--225, 2012.

\bibitem[Hesse and Gross(2014)]{hesse2014}
Janina Hesse and Thilo Gross.
\newblock Self-organized criticality as a fundamental property of neural
  systems.
\newblock \emph{Frontiers in systems neuroscience}, 8:\penalty0 166, 2014.

\bibitem[Neto et~al.(2017)Neto, de~Aguiar, Brum, and Bornholdt]{neto2017}
Joao~Pinheiro Neto, Marcus~AM de~Aguiar, Jos{\'e}~A Brum, and Stefan Bornholdt.
\newblock Inhibition as a determinant of activity and criticality in dynamical
  networks.
\newblock \emph{arXiv preprint arXiv:1712.08816}, 2017.

\bibitem[Keck et~al.(2017)Keck, Toyoizumi, Chen, Doiron, Feldman, Fox,
  Gerstner, Haydon, H{\"u}bener, Lee, et~al.]{keck2017}
Tara Keck, Taro Toyoizumi, Lu~Chen, Brent Doiron, Daniel~E Feldman, Kevin Fox,
  Wulfram Gerstner, Philip~G Haydon, Mark H{\"u}bener, Hey-Kyoung Lee, et~al.
\newblock Integrating hebbian and homeostatic plasticity: the current state of
  the field and future research directions.
\newblock \emph{Philosophical Transactions of the Royal Society B: Biological
  Sciences}, 372\penalty0 (1715):\penalty0 20160158, 2017.

\bibitem[Mediano and Shanahan(2017)]{mediano2017}
Pedro~AM Mediano and Murray Shanahan.
\newblock Balanced information storage and transfer in modular spiking neural
  networks.
\newblock \emph{arXiv preprint arXiv:1708.04392}, 2017.

\bibitem[Tax et~al.(2017)Tax, Mediano, and Shanahan]{tax2017}
Tycho Tax, Pedro~AM Mediano, and Murray Shanahan.
\newblock The partial information decomposition of generative neural network
  models.
\newblock \emph{Entropy}, 19\penalty0 (9):\penalty0 474, 2017.

\bibitem[Bill et~al.(2010)Bill, Schuch, Br{\"u}derle, Schemmel, Maass, and
  Meier]{bill2010}
Johannes Bill, Klaus Schuch, Daniel Br{\"u}derle, Johannes Schemmel, Wolfgang
  Maass, and Karlheinz Meier.
\newblock Compensating inhomogeneities of neuromorphic {VLSI} devices via
  short-term synaptic plasticity.
\newblock \emph{Frontiers in computational neuroscience}, 4:\penalty0 129,
  October 2010.

\bibitem[Wilting and Priesemann(2019)]{wilting2019}
J~Wilting and V~Priesemann.
\newblock 25 years of criticality in neuroscience--established results, open
  controversies, novel concepts.
\newblock \emph{arXiv preprint arXiv:1903.05129}, 2019.

\bibitem[Pfeil et~al.(2013)Pfeil, Gr{\"u}bl, Jeltsch, M\"uller, M\"uller,
  Petrovici, Schmuker, Br\"uderle, Schemmel, and Meier]{pfeil2013}
Thomas Pfeil, Andreas Gr{\"u}bl, Sebastian Jeltsch, Eric M\"uller, Paul
  M\"uller, Mihai~A. Petrovici, Michael Schmuker, Daniel Br\"uderle, Johannes
  Schemmel, and Karlheinz Meier.
\newblock Six networks on a universal neuromorphic computing substrate.
\newblock \emph{Frontiers in Neuroscience}, 7:\penalty0 11, 2013.

\bibitem[Br\"uderle et~al.(2011)Br\"uderle, Petrovici, Vogginger, Ehrlich,
  Pfeil, Millner, Gr{\"u}bl, Wendt, M\"uller, Schwartz, and
  et~al.]{bruederle2011}
Daniel Br\"uderle, Mihai~A. Petrovici, Bernhard Vogginger, Matthias Ehrlich,
  Thomas Pfeil, Sebastian Millner, Andreas Gr{\"u}bl, Karsten Wendt, Eric
  M\"uller, Marc-Olivier Schwartz, and et~al.
\newblock A comprehensive workflow for general-purpose neural modeling with
  highly configurable neuromorphic hardware systems.
\newblock \emph{Biological Cybernetics}, 104:\penalty0 263--296, 2011.

\bibitem[Neftci and Indiveri(2010)]{neftci2010}
E.~Neftci and G.~Indiveri.
\newblock A device mismatch compensation method for {VLSI} neural networks.
\newblock In \emph{2010 Biomedical Circuits and Systems Conference (BioCAS)},
  pages 262--265, Nov 2010.
\newblock \doi{10.1109/BIOCAS.2010.5709621}.

\bibitem[Neftci et~al.(2011)Neftci, Chicca, Indiveri, and Douglas]{neftci2011}
Emre Neftci, Elisabetta Chicca, Giacomo Indiveri, and Rodney Douglas.
\newblock A systematic method for configuring {VLSI} networks of spiking
  neurons.
\newblock \emph{Neural Computation}, 23\penalty0 (10):\penalty0 2457--2497,
  October 2011.
\newblock ISSN 0899-7667.

\bibitem[Morrison et~al.(2008)Morrison, Diesmann, and Gerstner]{morrison2008}
Abigail Morrison, Markus Diesmann, and Wulfram Gerstner.
\newblock Phenomenological models of synaptic plasticity based on spike timing.
\newblock \emph{Biological cybernetics}, 98\penalty0 (6):\penalty0 459--478,
  April 2008.

\bibitem[Goodman and Brette(2009)]{goodman2009}
Dan~FM Goodman and Romain Brette.
\newblock The brian simulator.
\newblock \emph{Frontiers in neuroscience}, 3:\penalty0 26, 2009.

\bibitem[Alstott et~al.(2014)Alstott, Bullmore, and Plenz]{alstott2014}
Jeff Alstott, Ed~Bullmore, and Dietmar Plenz.
\newblock powerlaw: a python package for analysis of heavy-tailed
  distributions.
\newblock \emph{PloS one}, 9\penalty0 (1):\penalty0 e85777, 2014.

\bibitem[Priesemann et~al.(2009)Priesemann, Munk, and Wibral]{priesemann2009}
Viola Priesemann, Matthias~HJ Munk, and Michael Wibral.
\newblock Subsampling effects in neuronal avalanche distributions recorded in
  vivo.
\newblock \emph{BMC neuroscience}, 10\penalty0 (1):\penalty0 40, 2009.

\bibitem[Priesemann et~al.(2014)Priesemann, Wibral, Valderrama, Pr{\"o}pper,
  Le~Van~Quyen, Geisel, Triesch, Nikoli{\'c}, and Munk]{priesemann2014}
Viola Priesemann, Michael Wibral, Mario Valderrama, Robert Pr{\"o}pper, Michel
  Le~Van~Quyen, Theo Geisel, Jochen Triesch, Danko Nikoli{\'c}, and Matthias~HJ
  Munk.
\newblock Spike avalanches in vivo suggest a driven, slightly subcritical brain
  state.
\newblock \emph{Frontiers in systems neuroscience}, 8, 2014.

\bibitem[Wei and Winnicki(1990)]{wei1990}
CZ~Wei and J~Winnicki.
\newblock Estimation of the means in the branching process with immigration.
\newblock \emph{The Annals of Statistics}, pages 1757--1773, 1990.

\bibitem[Lizier(2014)]{lizier2014}
Joseph~T Lizier.
\newblock Jidt: An information-theoretic toolkit for studying the dynamics of
  complex systems.
\newblock \emph{Frontiers in Robotics and AI}, 1:\penalty0 11, 2014.

\bibitem[Makkeh et~al.(2018)Makkeh, Theis, and Vicente]{makkeh2018}
Abdullah Makkeh, Dirk Theis, and Raul Vicente.
\newblock Broja-2pid: A robust estimator for bivariate partial information
  decomposition.
\newblock \emph{Entropy}, 20\penalty0 (4):\penalty0 271, 2018.

\end{thebibliography}
\providecommand{\noopsort}[1]{}\providecommand{\singleletter}[1]{#1}%

\end{document}